\documentclass[a4papper,12pt,feynmf]{article}
\usepackage{amssymb}
\usepackage{graphicx}
\usepackage{ifpdf}
\begin{document}
\def\ctr#1{\hfil $\,\,\,#1\,\,\,$ \hfil}
\def\tstrut{\vrule height 2.7ex depth 1.0ex width 0pt}
\def\mystrut{\vrule height 3.7ex depth 1.6ex width 0pt}
\def \inparg{\leftskip = 40pt\rightskip = 40pt}
\def \outparg{\leftskip = 0 pt\rightskip = 0pt}
\def\lf{16\pi^2}
\def\beqn{\begin{eqnarray}}
\def\eeqn{\end{eqnarray}}
\def\sy{supersymmetry}
\def\sic{supersymmetric}
\def\sa{supergravity}
\def\ssm{supersymmetric standard model}
\def\sm{standard model}
\def\ssb{spontaneous symmetry breaking}
\def\smgroup{$SU_3\otimes\ SU_2\otimes\ U_1$}
\def\app{{Acta Phys.\ Pol.\ }{\bf B}}
\def\anp{Ann.\ Phys.\ }
\def\cmp{Comm.\ Math.\ Phys.\ }
\def\fortphys{{Fort.\ Phys.\ }{\bf A}}
\def\ijmpa{{Int.\ J.\ Mod.\ Phys.\ }{\bf A}}
\def\jetp{JETP\ }
\def\jetpl{JETP Lett.\ }
\def\jmp{J.\ Math.\ Phys.\ }
\def\mpla{{Mod.\ Phys.\ Lett.\ }{\bf A}}
\def\nc{Nuovo Cimento\ }
\def\npb{{Nucl.\ Phys.\ }{\bf B}}
\def\physrep{Phys.\ Reports\ }
\def\plb{{Phys.\ Lett.\ }{\bf B}}
\def\pnas{Proc.\ Natl.\ Acad.\ Sci.\ (U.S.)\ }
\def\pr{Phys.\ Rev.\ }
\def\prd{{Phys.\ Rev.\ }{\bf D}}
\def\prl{Phys.\ Rev.\ Lett.\ }
\def\ptp{Prog.\ Th.\ Phys.\ }
\def\sjnp{Sov.\ J.\ Nucl.\ Phys.\ }
\def\tmp{Theor.\ Math.\ Phys.\ }
\def\pw{Part.\ World\ }
\def\zpc{Z.\ Phys.\ {\bf C}}
\def\pa{\partial}

\vskip .3in {\textbf{\Large{ Comments on the role of field redefinition on  renormalisation of   $N=\frac{1}{2}$ supersymmetric pure gauge theory }}}
\medskip
\vskip .3in \centerline{\bf   A. F. Kord $^{1,2}$,
M.~Haddadi Moghaddam $^{1}$ }
\bigskip
{\small{\it {$1$: Department of Physics, Hakim Sabzevari
University (HSU), P.O.Box 397, Sabzevar, Iran

 $2$: Institute for Research  in Fundamental Sciences(IPM),P.O. Box 19395-5531, Tehran, Iran }}}

\small{\it\center{E-mail: afarzaneh@hsu.ac.ir}} \vskip .3in
\begin{abstract}
We study one loop corrections to $N=\frac{1}{2}$ supersymmetric $SU(N)\times U(1)$ pure gauge theory. We calculate   divergent contributions of the 1PI
graphs contain the non-anti-commutative parameter $C$ up to one loop corrections. We find the  disagreement between component formalism and  superspace formalism is because of  the field redefinition in component case. We modify gaugino      field redefinition and  lagrangian. We show  extra terms of lagrangian have been generated  by $\lambda$ redefinition and are necessary for the renormalisation of the theory. Finally  we prove  $N = \frac{1}{2}$ supersymmetric gauge theory is renormalisable up one loop corrections using standard method of renormalisation
\end{abstract}

\section{Introduction}
Theories defined on non-anti-commutative superspace have been
studied extensively during last ten years ~\cite{a1, a2}.
Superspace in such  non-anti-commutative theories is a superspace
whose fermionic supercoordinates are not anticommutative. One
could construct a field theory in non-anti-commutative superspace
in terms of superfields with the star-product where lagrangian is
deformed from the original theory by the non-anti-commutative
parameters.

Recently some renormalisability   aspects of  the
non-anti-commutative field theories have been studies. It has
been shown non-anti-commutative field theories are not
power-counting renormalisable; however it has been discussed that
they could be renormalisable if some additional terms have been
added to the lagrangian in order to divergences to all
orders~\cite{a3}-\cite{a8}. The renormalisability of
non-anti-commutative versions of the Wess-Zumino model has been
discussed ~\cite{a3,a4}, with explicit computations up to two
loops ~\cite{a5}. The renormalisability of non-anti-commutative
gauge theory with $N =\frac{1}{2}$ supersymmetry has been studied
in~\cite{a6,a8}. The authors in ~\cite{a6} show that the theory is
renormalisable to all order of perturbation theory. The
conditions of the  renormalisability of non-anticommutative (NAC) field
theories have been studied with  explicit computations up one and
two loops~\cite{a9}-\cite{a15}.

The renormalisability of supersymmetric gauge field theories has been discussed in WZ gauge ~\cite{a6,a7}. The explicit  one loop corrections in  component formalism have been done in ~\cite{a9}-\cite{a11}.
The authors in~\cite{a10,a11} have claimed  the precise form of the
lagrangian is not  preserved by renormalisation.  They have  shown  by explicit
calculation that there are problems with assumption of gauge invariance   which is required to rule out some
classes of divergent structure in non-anti-commutative theory.From their calculation, one can see even at one loop
divergent non-gauge-invariant terms are generated. In order to remove the
non-gauge-invariant terms and restore gauge invariance at one loops they introduce  a one loop divergent field
redefinition in the case of pure $N =\frac{1}{2}$ supersymmetry (i.e. no chiral matter).

On the other hand, the authors in~\cite{a12,a13} have started from superspace formalism and  discussed renormalisability and  supergauge  invariance. They proved that the field redefinition is  not necessary and the original  effective action is not only gauge but also supergauge invariant up one loop corrections. The disagreement between two approaches   put a big question mark which approaches we should relay on in $N=\frac{1}{2}$ supersymmetric gauge theory.

In this paper we investigate the renormalisability of $N
=\frac{1}{2}$ supersymmetric  pure  gauge theory at one-loop
perturbative corrections in component formalism.  We shall show $N =\frac{1}{2}$
supersymmetric gauge theory is renormalisable in a usual manner
without any needs  for   field redefinition (there is not
theoretical justification or interpretation for the field
redefinition as mentioned by authors ~\cite{a10})  which leads to
the lagrangian change. Therefore we shall prove two approaches lead to the same conclusion.

The paper is organized as follows: First we briefly review NAC supersymmetric gauge theories and their lagrangian. Then an explicit one-loop calculation of the
three and four-point functions of the theory in the C-deformed
sector is carried out to calculate the corrections. We show some
anomaly  terms  appears in the 1PI functions which spoil
the renormalisability of theory. Next we introduce extra terms to the original lagrangian in order to renormalise NAC pure gauge supersymmetric theory and calculate   corrections  which come from these new terms. Finally we discus  the source of the extra lagrangian, and  show that these new terms have hidden because of  the component $\lambda$ redefinition  ~\cite{a1,a20}), so in order to reproduce them one should reverse gaugino  field  redefinition.

\section{The pure gauge supersummetric action of NAC gauge theory}

The original non-anticommutative theory defined in superfields
appears to require a U(N) gauge group. Here, at first we would like to
consider U(N) gauge theory for non-(anti)commutative (NAC) superspace. The action for an
N = 1/2 supersymmetric U(N) pure gauge theory is given by:
\begin{eqnarray}
S&=&\int
d^4x\Big[Tr\{-\frac{1}{2}F^{\mu\nu}F_{\mu\nu}-2i{\bar\lambda}{\bar\sigma}^\mu{(D_\mu\lambda)}+
D^2\}\nonumber\\&&-2igC^{\mu\nu}Tr\{F_{\mu\nu}{\bar\lambda}{\bar\lambda}\}
+g^2\mid C\mid^2Tr\{({\bar\lambda}{\bar\lambda})^2\}\Big],
\end{eqnarray}
where
\begin{eqnarray}
F_{\mu\nu}&=&\partial_\mu A_\nu-\partial_\nu A_\mu+ig[A_\mu , A_\nu],\nonumber\\
D_\mu\lambda&=&\partial_\mu\lambda+ig[A_\mu , \lambda],
\end{eqnarray}
and
\begin{eqnarray}
A_\mu=A_\mu^AR^A,\ \lambda=\lambda^AR^A,\ D=D^AR^A,
\end{eqnarray}
Corresponding to any index $a$ for SU(N) we introduce the index
$A = (0, a)$, so that $A$ runs from 0 to $N^2 - 1$. with $R^A$
being the group matrices for U(N) in the fundamental
representation. These satisfy
\begin{eqnarray}
[R^A , R^B]&=&if^{ABC}R^C, \ \{R^A , R^B\}=d^{ABC}R^C,
\end{eqnarray}
where $f^{ABC}$ is completely antisymmetric, $f^{abc}$ is the same
as for SU(N) and $f^{0bc} = 0$, while $d^{ABC}$ is completely
symmetric; $d^{abc}$ is the same as for SU(N), $d^{0bc} = \sqrt{
2/N}\delta^{bc}, d^{00c} = 0$ and $d^{000} = \sqrt{2/N}$. In
particular, $R^0 = \sqrt{ \frac{1}{2N}} 1$ . We have also
\begin{eqnarray}
Tr\{R^AR^B\}=\frac{1}{2}\delta^{AB}
\end{eqnarray}
The following identities hold in U(N) group and will be extensively used below
\begin{eqnarray}
&&f^{ABL}f^{LCD}+f^{ACL}f^{LDB}+f^{ADL}f^{LBC}=0,\\&&
f^{ABL}d^{LCD}+f^{ACL}d^{LDB}+f^{ADL}d^{LBC}=0,\\&&
f^{ADL}f^{LBC}=d^{ABL}d^{LCD}-d^{ACL}d^{LDB},\\&&
f^{IAJ}f^{JBK}f^{KCI}=-\frac{N}{2}f^{ABC},\\&&
d^{IAJ}f^{JBK}f^{KCI}=-\frac{N}{2}d^{ABC}d^A c^B c^C.
\end{eqnarray}
Where $d^A=1+\delta^{0A}$ , $c^A=1-\delta^{0A}$.

Upon substituting  the above relations in eq. (1), we obtain the action in
the U(N) case in the form:
\begin{eqnarray}
S&=&\int d^4x\Big[-\frac{1}{4}F^{\mu\nu A
}F_{\mu\nu}^A-i{\bar\lambda}^A{\bar\sigma}^\mu{(D_\mu\lambda)}^A+\frac{1}{2}
D^AD^A\nonumber\\&&-\frac{1}{2}igd^{ABC}C^{\mu\nu}F_{\mu\nu}^A{\bar\lambda}^B{\bar\lambda}^C
+\frac{1}{8}g^2 d^{ABE}d^{CDE}\mid
C\mid^2({\bar\lambda}^A{\bar\lambda}^B)({\bar\lambda}^C{\bar\lambda}^D)\Big]\nonumber\\,
\end{eqnarray}
With gauge coupling $g$, gauge field $A_\mu$ and gaugino
$\lambda$.

Beside, definition for $F_{\mu\nu}$ and $D_\mu\lambda^a$ are given by:
\begin{eqnarray}
F_{\mu\nu}^A&=&\pa_\mu A_\nu^A-\pa_\nu A_\mu^A-gf^{ABC}A_\mu^BA_\nu^C,\nonumber\\
D_\mu\lambda^A&=&\pa_\mu\lambda^A-gf^{ABC}A_\mu^B\lambda^C,
\end{eqnarray}
$C^{\mu\nu}$ is related to the non-anti-commutativity parameter
$C^{\alpha\beta}$ by:
\begin{eqnarray}
C^{\mu\nu}&=&C^{\alpha\beta}\epsilon_{\beta\gamma}\sigma_\alpha^{\mu\nu\
\gamma}
\end{eqnarray}
also, we have:
\begin{eqnarray}
C^{\alpha\beta}&=&\frac{1}{2}\epsilon^{\alpha\gamma}\sigma_\gamma^{\mu\nu\
\beta} C_{\mu\nu},
\end{eqnarray}
where
\begin{eqnarray}
\sigma_\alpha^{\mu\nu\
\beta}=\frac{1}{4}({\sigma^\mu}_{\alpha\dot\rho}{\bar\sigma}^
{\nu\dot\rho\beta}-{\sigma^\nu}_{\alpha\dot\rho}{\bar\sigma}^
{\mu\dot\rho\beta}),\\
{\bar\sigma}^{\mu\nu\dot\alpha}_{\dot\beta}=\frac{1}{4}
({\bar\sigma}^{\mu\dot\alpha\rho}{\sigma^\nu}_{\rho\dot\beta}-
{\bar\sigma}^{\nu\dot\alpha\rho}{\sigma^\mu}_{\rho\dot\beta}).
\end{eqnarray}
The useful identity  is:
\begin{eqnarray}
\mid C\mid^2&=&C^{\mu\nu}C_{\mu\nu}.
\end{eqnarray}
There are some properties of C in App A.  In above Eqs.,
$C^{\alpha\beta}$ is the non-anti-commutativity parameter, and
our conventions are consistent  with ref~\cite{a1}.
The action
for pure $N = 1/ 2$ supersymmetric gauge theory (Eq.~11) is
invariant under the standard $U(N)$ gauge transformations and
the $N = 1/2$ supersymmetry transformations.
The standard $U(N)$  version of the NAC gauge theory is not renormalisable~\cite{a1}. Therefore; we would like to present a $SU(N)\times U(1)$ lagrangian which has $N=\frac{1}{2}$ supersymmetric properties, so we introduce the following action:
\begin{eqnarray}
S&=&\int d^4x\Big[-\frac{1}{4}F^{\mu\nu A
}F_{\mu\nu}^A-i{\bar\lambda}^A{\bar\sigma}^\mu{(D_\mu\lambda)}^A+\frac{1}{2}
D^AD^A\nonumber\\&&-\frac{1}{2}i\gamma^{ABC}d^{ABC}C^{\mu\nu}F_{\mu\nu}^A{\bar\lambda}^B{\bar\lambda}^C
+\frac{1}{8}\gamma^{ABCDE} d^{ABE}d^{CDE}\mid
C\mid^2({\bar\lambda}^A{\bar\lambda}^B)({\bar\lambda}^C{\bar\lambda}^D)\Big]\nonumber\\,
\end{eqnarray}
One beauty of the above equation is one could easily switch between the original $U(N)$ theory and $SU(N) \times U(1)$ theory. In our work we
 define $\gamma^{abcde}=\gamma_0,\ \gamma^{abcd0}=\gamma_1,\ \gamma^{0b0de}=\gamma^{a0c0e}=
\gamma^{0bc0e}=\gamma^{a00de}= \gamma_2$. Indeed we give  them in terms of $g$ and $g_0$. They are given by:
\begin{eqnarray}
&&\gamma^{abc}=g,\ \gamma^{ab0}=\gamma^{a0b}=g_0,\ \gamma^{0ab}=\frac{g^2}{g_0}\\&&
\gamma_0=g^2,\ \gamma_1=(\frac{g^2}{g_0})^2,\ \gamma_2=g_0^2 h
\end{eqnarray}
Where $h=1$. The above action is similar to the $SU(N)\times U(1)$ action in ref~\cite{a11}.

 The N=1/2 supersymmetry transformation is:
\begin{eqnarray}
&&\delta A_\mu^A=-i{\bar\lambda}^A{\bar\sigma}_\mu\epsilon ~,\\&&
\delta\lambda_\alpha^A=i\epsilon_\alpha
D^A+{(\sigma^{\mu\nu}\epsilon)}_\alpha[F_{\mu\nu}^A+\frac{1}{2}iC_{\mu\nu}\gamma^{ABC}d^{ABC}{\bar\lambda}^B{\bar\lambda}^C],\
\delta {\bar\lambda_{\dot\alpha}}^A=0~,\\&& \delta D^A=-\epsilon\sigma^\mu
D_\mu {\bar\lambda}^A.
\end{eqnarray}

 \section{One-loop  corrections}

 In our calculation, we use standard gauge fixing term
 \begin{eqnarray}
 S_{gf}&=\frac{1}{2\alpha} &\int d^4x(\partial.A)^2
 \end{eqnarray}
 and consider the  Feynman rules in the super-Fermi-Feynman gauge$(\alpha=1)$.

In this section we first review  the one-loop perturbative corrections to the undeformed
$N = 1$ part of the theory. It has been shown that the quantum
corrections of $N = 1$ part of the theory are not affected by
$C$-deformation~\cite{a9,a10}. Therefore; gauge field and gaugino
anomalous dimensions and gauge $\beta$-functions are the same  as
those in the ordinary $N = 1$ case.
 The $C$-independent part of the bare action can be written as:
\begin{eqnarray}
S_{C=0}=\int d^4x\Big[-\frac{1}{4}F^{\mu\nu\
A}F_{\mu\nu}^A-i{\bar\lambda}^A{\bar\sigma}^\mu\pa_\mu\lambda^A
+igf^{ABC}{\bar\lambda}^A{\bar\sigma}^\mu
A_\mu^B\lambda^C+\frac{1}{2}D^AD^A\Big]
\end{eqnarray}
The $C$-independent part of the action is renormalisable if one
introduce bare fields and couplings according to:
\begin{eqnarray}
 A_{B\mu}=Z_A^{\frac{1}{2}}A_{\mu},\ \  \lambda_B=Z_{\lambda}^{\frac{1}{2}}\lambda,\ \   g_B=Z_gg,
\end{eqnarray}
that $Z_A, Z_\lambda$ and $Z_g$ are known as a renormalisation
constants. Also one can define:
\begin{eqnarray}
\delta_1=Z_A-1, \delta_2=Z_\lambda-1,\ \
\delta_3=Z_gZ_A^{\frac{1}{2}}Z_\lambda-1,
\end{eqnarray}
finally, one should add  the following counter terms to the
lagrangian of  theory in order to renormalise  theory:
\begin{eqnarray}
L_{counter-terms}=
-\frac{1}{4}\delta_1F^{\mu\nu\
A}F_{\mu\nu}^A-i\delta_2{\bar\lambda}^A{\bar\sigma}^\mu\pa_\mu\lambda^A
+\delta_3igf^{ABC}{\bar\lambda}^A{\bar\sigma}^\mu
A_\mu^B\lambda^C,
\end{eqnarray}
where,
\begin{eqnarray}
Z_A=1+2NL,\ \ \ Z_\lambda=1-2NL, \ \ \ \ Z_g=1-3NL,
\end{eqnarray}
and $L$ is given by:
\begin{eqnarray}
L=\frac{g^2}{16\pi^2\varepsilon}.
\end{eqnarray}
Here $\varepsilon= 4-D$ is the regulator.

(We have given here the renormalisation constants corresponding to the $SU(N)$ sector
of the $U(N)$ theory; those for the $U(1)$ sector, namely $Z_{\lambda^0}, Z_{A^0}, Z_{g_0}$
are given by ommiting the terms in $N$ and replacing $g$ by $g_0$. )

 \subsection{One-loop $C$ deformed Corrections}
In this part we will present on-loop graphs contributing to
the new terms arising from $C$-deformed part of the action. The
one-loop one-particle-irreducible (1PI) graphs of the  $C$-deform
$A\bar{\lambda}\bar{\lambda}$ three point functions are depicted
in Figs(2). Using Feynman rules one could compute the divergent
contributions  from the graphs. As an example we calculate the
one loop corrections to fig(2-a). It is given by:
\begin{figure}[ht]
\centerline{\includegraphics[width=8cm]{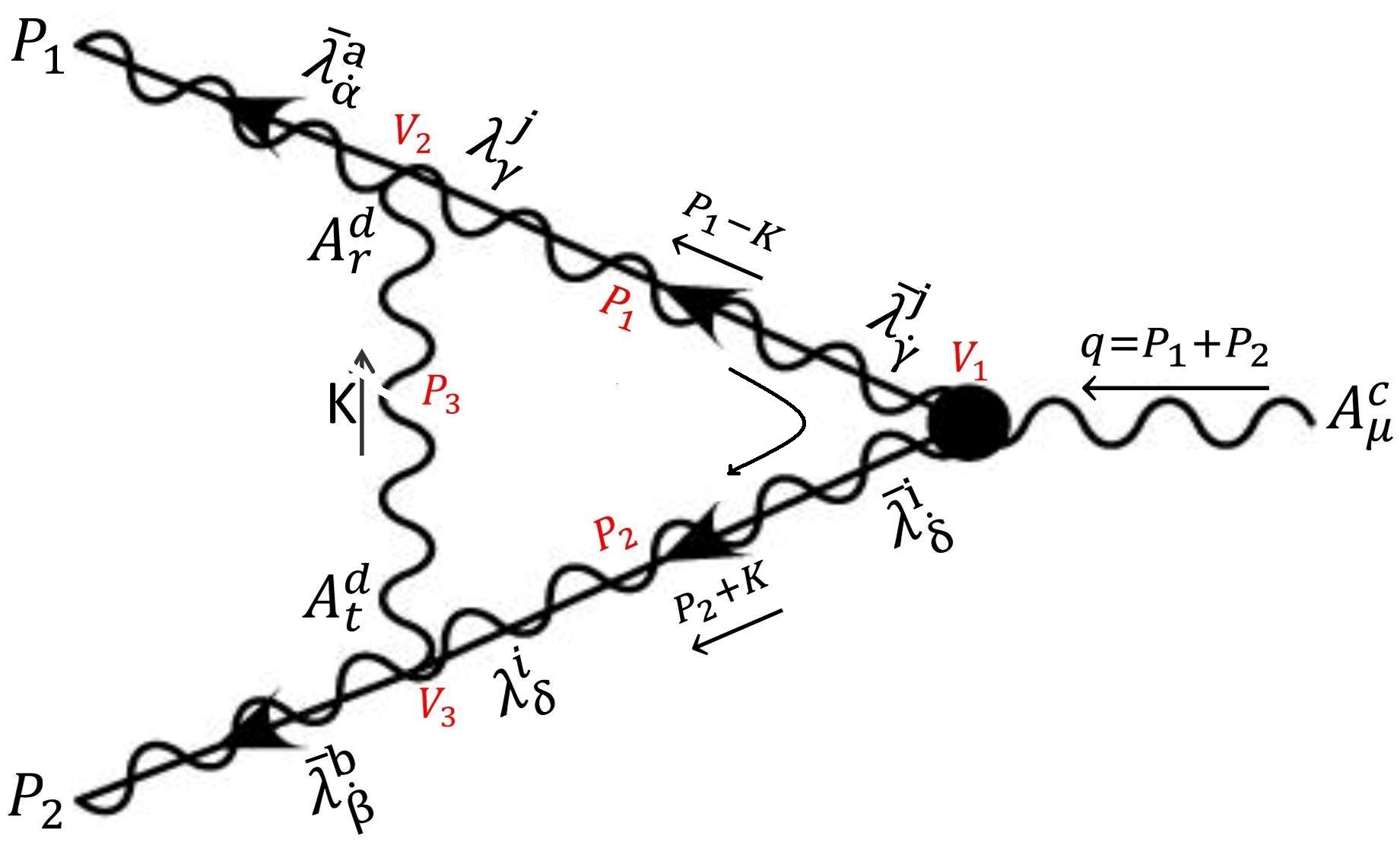}}
\caption{\label{fig1} The diagram  (2-a).}
\end{figure}
\begin{eqnarray}
\Gamma_a^{\mu\dot\alpha\dot\beta}&=&-g^2\gamma^{AJI}d^{AJI}f^{BDJ}f^{CDI}
C^{\mu\nu}\epsilon^{\dot\gamma\dot\delta}(p_1+p_2)_\nu
 \bar\sigma^{r\
\dot\alpha\gamma}\bar\sigma^{t\
\dot\beta\delta}\sigma^m_{\gamma\dot\gamma}\sigma^n_{\delta\dot\delta}\
g_{rt}\nonumber\\&&
\times\int\frac{d^dk}{(2\pi)^d}\frac{(p_1-k)_k(p_2+k)_\lambda}{k^2(p_1-k)^2(p_2+k)^2}
\end{eqnarray}
Using  Feynman parameter in App. B:
\begin{eqnarray}
\frac{1}{abc}=2\int_0^1dx\int_0^1dy\frac{x}{[axy+bx(1-y)+c(1-x)]^3},
\end{eqnarray}
 we can simplify denominator of Eq. (31)
\begin{eqnarray}
&&\frac{1}{k^2(p_1-k)^2(p_2+k)^2}=2\int_0^1dx\int_0^1dy\nonumber\\&&
\times\frac{x}{[k^2+2k.[p_1x(y-1)+p_2(1-x)]+p_1^2x(1-y)+p_2^2(1-x)]^3}.
\end{eqnarray}
By changing variables to
$$k'=k+p_1x(y-1)+p_2(1-x),$$
the denominator of integral in Eq. (33) is given by:
$$\frac{1}{k^2(p_1-k)^2(p_2+k)^2}=2\int_0^1dx\int_0^1dy\frac{x}{[k^2-\Delta]^3}$$
where
$$\Delta=[p_1x(2y-1)+2p_2(1-x)]^2$$
so, the integral of Eq. (32) is given by:
\begin{eqnarray}
2\int_0^1dx\int_0^1dy\int\frac{d^dk}{(2\pi)^d}\frac{x(p_1-k)_k(p_2+k)_\lambda}{[k^2-\Delta]^3}
\end{eqnarray}
then we arrive:
\begin{eqnarray}
2\int_0^1dx\int_0^1dy\int\frac{d^dk}{(2\pi)^d}\frac{-k_kk_\lambda}{[k^2-\Delta]^3}
=\frac{-ig_{k\lambda}}{32\pi^2\varepsilon}
\end{eqnarray}
we  finally have for Eq. (31):
\begin{eqnarray}
\Gamma_a^{\mu\dot\alpha\dot\beta}&=&
4iNL\gamma^{ABC}d^{ABC}d^Ac^Bc^C\epsilon^{\dot\alpha\dot\beta}C^{\mu\rho}(p_1+p_2)_\rho
\end{eqnarray}
 Moreover, as it be seen in Fig. (2-a), we
have momentum - energy conserving in the loop:
$$q_\nu={(p_1+p_2)}_\nu$$

 The divergent contributions up to one loop correction to  diagrams in Fig. 2 are given by:
  \begin{eqnarray}\Gamma_{2-a}^{(1)\mu\dot\alpha\dot\beta}&=&4iNL\gamma^{ABC}d^{ABC}d^Ac^Bc^C\epsilon^{\dot\alpha\dot\beta}C^{\mu\nu}q_\nu
\nonumber\\
\Gamma_{2-b}^{(1)\mu\dot\alpha\dot\beta}&=&iNL\gamma^{ABC}d^{ABC}c^Ad^Bc^C[\frac{1}{2}\epsilon^{\dot\alpha\dot\beta}
C^{\mu\nu}q_\nu
+\frac{1}{3}{(Y^{\mu\nu})}^{\dot\alpha\dot\beta}(p_1-p_2)_\nu]\nonumber\\&&+one\ permutation\nonumber\\
\Gamma_{2-c}^{(1)\mu\dot\alpha\dot\beta}&=&\frac{1}{4}iNL\gamma^{ABC}d^{ABC}c^Ad^Bc^C[\epsilon^{\dot\alpha\dot\beta}C^{\mu\nu}(4p_1+5p_2)_\nu
-\frac{2}{3}{(Y^{\mu\nu})}^{\dot\alpha\dot\beta}(2p_1+7p_2)_\nu]\nonumber\\&&+one\ permutation\nonumber\\
\Gamma_{2-d}^{(1)\mu\dot\alpha\dot\beta}&=&-3iNL\gamma^{ABC}d^{ABC}c^A\epsilon^{\dot\alpha\dot\beta}C^{\mu\nu}q_\nu\nonumber\\
\Gamma_{2-e}^{(1)\mu\dot\alpha\dot\beta}&=&-\frac{1}{2}iNL\gamma^{ABC}d^{ABC}c^Ad^Bc^C[\epsilon^{\dot\alpha\dot\beta}C^{\mu\nu}
-2{(Y^{\mu\nu})}^{\dot\alpha\dot\beta}]p_{2\nu}\nonumber\\&&+one\ permutation
\end{eqnarray}
$Y^{\mu\nu}$is been  defined,
\begin{eqnarray}
&&({Y^{\mu\nu}})^{\dot\alpha\dot\beta}=\epsilon^{\dot\alpha\dot\theta}C^{\mu\rho}g_{\rho\lambda}({{\bar\sigma}^{\lambda\nu}})^{\dot\beta}_{{}{\dot\theta}}
\end{eqnarray}
Where tensor $Y^{\mu\nu}$ is symmetric respect to both Lorentz and spinor indices  and tensor $C^{\mu\nu}$ is
anti-symmetric.
In our computation  permutations has taken into account by changing the position of $C$ as well as symmetry factors. Adding the different divergent contributions from the diagrams of  fig~2 corresponding to
different $U(1)$ and $SU(N)$  parts, we have:

\begin{eqnarray}
\Sigma_{i=a}^{e}\Gamma_{2-i}^{(1)\mu\dot\alpha\dot\beta}&=&\frac{15}{4}iNL\gamma^{abc}d^{abc}\epsilon^{\dot\alpha\dot\beta}C^{\mu\nu}q_\nu+8iNL\gamma^{0bc}d^{0bc}\epsilon^{\dot\alpha\dot\beta}C^{\mu\nu}q_\nu\nonumber\\&&
-\frac{1}{2}iNL\gamma^{a0c}d^{a0c}\epsilon^{\dot\alpha\dot\beta}C^{\mu\nu}p_{2\nu}-\frac{1}{2}iNL\gamma^{ab0}d^{ab0}\epsilon^{\dot\alpha\dot\beta}C^{\mu\nu}p_{1\nu}\nonumber\\&&
+\frac{1}{2}iNL\gamma^{abc}d^{abc}(Y^{\mu\nu})^{\dot\alpha\dot\beta}(p_1-p_2)_\nu\nonumber\\&&-iNL\gamma^{a0c}d^{a0c}(Y^{\mu\nu})^{\dot\alpha\dot\beta}p_{2\nu}+iNL\gamma^{ab0}d^{ab0}(Y^{\mu\nu})^{\dot\alpha\dot\beta}p_{1\nu}
\end{eqnarray}

Let us now continue with the relevant  diagrams containing only
$C$-deformed vertex and contributing to the four point functions
(Fig.~3 and Fig.~4 ). Using the Feynman rules, and  considering
all  permutations between the same fields, the final result for
1PI graphs of Fig.~3 are given by:
\begin{eqnarray}
\Gamma_{3-a}^{(1)\mu\nu\dot\alpha\dot\beta}&=&NLg[\gamma^{EAB}d^{ABE}f^{CDE}-\gamma^{CDE}f^{ABE}d^{CDE}
]d^Ac^Bc^Cc^D\nonumber\\&&[\frac{1}{4}\epsilon^{\dot\alpha\dot\beta}C^{\mu\nu}-\frac{1}{6}{(Y^{\mu\nu})}^{\dot\alpha\dot\beta}]+Three\ permutations\nonumber\\
\Gamma_{3-b}^{(1)\mu\nu\dot\alpha\dot\beta}&=&NLg[\gamma^{EAC}d^{ACE}f^{BDE}+\gamma^{BDE}f^{ACE}d^{BDE}
]d^Ac^Bc^Cc^D\nonumber\\&&[\frac{1}{8}\epsilon^{\dot\alpha\dot\beta}C^{\mu\nu}-\frac{1}{4}{(Y^{\mu\nu})}^{\dot\alpha\dot\beta}]+One\ permutations\nonumber\\
\Gamma_{3-c}^{(1)\mu\nu\dot\alpha\dot\beta}&=&2NLg\gamma^{EAB}d^{ABE}f^{CDE}c^Ac^B\epsilon^{\dot\alpha\dot\beta}C^{\mu\nu}\nonumber\\
\Gamma_{3-d}^{(1)\mu\nu\dot\alpha\dot\beta}&=&NLg[\gamma^{EAD}d^{ADE}f^{BCE}+\gamma^{BCE}f^{ADE}d^{BCE}
]d^Ac^Bc^Cc^D\nonumber\\&&[\frac{1}{8}\epsilon^{\dot\alpha\dot\beta}C^{\mu\nu}-\frac{5}{12}{(Y^{\mu\nu})}^{\dot\alpha\dot\beta}]+One\ permutations\nonumber\\
\Gamma_{3-e}^{(1)\mu\nu\dot\alpha\dot\beta}&=&-NLg[\gamma^{EAB}d^{ABE}f^{CDE}-\gamma^{CDE}f^{ABE}d^{CDE}
]d^Ac^Bc^Cc^D\nonumber\\&&[\frac{1}{4}\epsilon^{\dot\alpha\dot\beta}C^{\mu\nu}+\frac{7}{12}{(Y^{\mu\nu})}^{\dot\alpha\dot\beta}]+Three\ permutations\nonumber\\
\Gamma_{3-f}^{(1)\mu\nu\dot\alpha\dot\beta}&=&NLg[\gamma^{EAB}d^{ABE}f^{CDE}+\gamma^{CDE}f^{ABE}d^{CDE}
]d^Ac^Bc^Cc^D\nonumber\\&&[\frac{3}{16}\epsilon^{\dot\alpha\dot\beta}C^{\mu\nu}-\frac{3}{8}{(Y^{\mu\nu})}^{\dot\alpha\dot\beta}]+One\ permutations\nonumber\\
\Gamma_{3-g}^{(1)\mu\nu\dot\alpha\dot\beta}&=&NLg[\frac{3}{8}\gamma^{EAB}d^{ABE}f^{CDE}d^Ac^B\epsilon^{\dot\alpha\dot\beta}C^{\mu\nu}+\frac{1}{4}\gamma^{CDE}f^{ABE}d^{CDE}c^Cc^D{(Y^{\mu\nu})}^{\dot\alpha\dot\beta}
]\nonumber\\&&+One\ permutations\nonumber\\
\Gamma_{3-h}^{(1)\mu\nu\dot\alpha\dot\beta}&=&-\frac{3}{2}NLg\gamma^{EAB}d^{ABE}f^{CDE}\epsilon^{\dot\alpha\dot\beta}C^{\mu\nu}\nonumber\\
\Gamma_{3-i}^{(1)\mu\nu\dot\alpha\dot\beta}&=&\frac{3}{4}NLg\gamma^{EAB}d^{ABE}f^{CDE}\epsilon^{\dot\alpha\dot\beta}C^{\mu\nu}
\end{eqnarray}

Considering all diagrams the final result for Fig.~3 is given by:
\begin{eqnarray}
\Sigma_{i=a}^{i}\Gamma_{3-i}^{(1)\mu\nu\dot\alpha\dot\beta}&=&\frac{11}{4}NLg\gamma^{eab} d^{abe}f^{cde}\epsilon^{\dot\alpha\dot\beta}C^{\mu\nu}+Ng\gamma^{ea0} d^{a0e}f^{cde}\epsilon^{\dot\alpha\dot\beta}C^{\mu\nu}\nonumber\\&&
+\frac{1}{2}NLg\gamma^{ecd}d^{cde}f^{abe}{(Y^{\mu\nu})}^{\dot\alpha\dot\beta}
\end{eqnarray}

Finally, The final divergence contributions( Fig.~4) which come from the
last term  containing $\mid C\mid^2 (\bar\lambda\bar\lambda)^2$  are given by:
\begin{eqnarray}
\Gamma_{4-a}^{(1)\dot\alpha\dot\beta\dot\delta\dot\gamma}&=&iL\epsilon^{\dot\alpha\dot\beta}\epsilon^{\dot\delta\dot\gamma}
\mid C\mid^2[N\gamma_0d^{abe}d^{cde}+2\gamma_0d^{abcd}+4\gamma_1]\nonumber\\
\Gamma_{4-b}^{(1)\dot\alpha\dot\beta\dot\delta\dot\gamma}&=&iL\epsilon^{\dot\alpha\dot\beta}\epsilon^{\dot\delta\dot\gamma}
\mid C\mid^2[\frac{N}{2}\gamma_0d^{abe}d^{cde}-2\gamma_2]\nonumber\\
\Gamma_{4-c}^{(1)\dot\alpha\dot\beta\dot\delta\dot\gamma}&=&iL\epsilon^{\dot\alpha\dot\beta}\epsilon^{\dot\delta\dot\gamma}
\mid C\mid^2[-\frac{N}{2}\gamma_0d^{abe}d^{cde}+2\gamma_2]\nonumber\\
\Gamma_{4-d}^{(1)\dot\alpha\dot\beta\dot\delta\dot\gamma}&=&iL\epsilon^{\dot\alpha\dot\beta}\epsilon^{\dot\delta\dot\gamma}
\mid C\mid^2[-2\gamma_0 d^{abcd}+\frac{1}{3}(\tilde d^{abcd}-\tilde d^{acdb})+3\gamma_2]
\end{eqnarray}
Note that we have considered all permutations between the same
fields and changing the position of $C$ and $\mid C \mid^2$, and adds all divergences come from Fig. 4. The final
result for 1PI graphs of Fig. 4   is given by:
\begin{eqnarray}
\Sigma_{i=a}^{d}\Gamma_{4-i}^{(1)\dot\alpha\dot\beta\dot\delta\dot\gamma}&=&\frac{5}{4}iNL\gamma_0\epsilon^{\dot\alpha\dot\beta}\epsilon^{\dot\delta\dot\gamma}d^{abe}d^{cde}
\mid  C\mid^2+4iL\gamma_1\epsilon^{\dot\alpha\dot\beta}\epsilon^{\dot\delta\dot\gamma}
\mid C\mid^2\nonumber\\&&
+3iL\gamma_2\epsilon^{\dot\alpha\dot\beta}\epsilon^{\dot\delta\dot\gamma}\mid C\mid^2
\end{eqnarray}
In order to renormalise the theory and  remove the divergences one should rescale and redefine   coupling constants. This
procedure is equivalent to introduce of some counter-terms to the
lagrangian. Therefore,  we introduce bare couplings according to:
 \begin{eqnarray}
&& C^{\mu\nu}_B=Z_C C^{\mu\nu},\ \   \mid
C\mid^2_B=Z_{\mid C\mid^2}\mid C\mid^2,\\&&
\gamma^{ABC}_B=Z_{\gamma_{ABC}}\gamma^{ABC},\ \
\gamma^{ABCDE}_B=Z_{\gamma_{ABCDE}}\gamma^{ABCDE},
\end{eqnarray}

However in the language of counter-terms, a theory is
renormalisable if the counter-terms are of the same form as those
appearing in the original lagrangian( these counter terms are
required to cancel the divergences). If we look at three and four
point functions we see
 that the anomaly term
$$\gamma^{ABC}d^{ABC}{(Y^{\mu\nu})}^{\dot\alpha\dot\beta},\ \ g\gamma^{ECD}d^{CDE}f^{ABE}
{(Y^{\mu\nu})}^{\dot\alpha\dot\beta}$$
 called  $Y$ term is  problematic because we can not add some kinds of counter-terms which cancel $Y$ term, or one can  say these terms spoil renormalisation.
 In the next section we add some extra lagrangian to the theory and prove that the NAC pure gauge theory is renormalisable.

\section{Renormalization of $N=\frac{1}{2}$ deformed lagrangian}
In this section we shall renormalise  the theory and  remove the
divergences. In order to renormalise the theory we should add  extra term ($L_{Extra}$) to the original lagrangian. The extra lagrangian is considered  as follow:
\begin{eqnarray}
L_{Extra}&=&+\frac{i}{16}
d^{ABC}\kappa^{BAC}C^{\mu\nu}(\partial_\mu A^A_\nu-\partial_\nu
A^A_\mu)\bar\lambda^B\bar\lambda^C\nonumber\\&&-\frac{i}{4}
g\kappa^{EDB}d^{BDE}f^{ACE} C^{\mu\nu} A_\mu^C A_\nu^D
\bar\lambda^A \bar\lambda^B\nonumber\\&&+\frac{i}{4}
\kappa^{BAC}d^{ABC}A_\mu^A(\partial_\nu
{\bar\lambda}^BY^{\mu\nu}{\bar\lambda}^C-{\bar\lambda}^BY^{\mu\nu}\partial_\nu{\bar\lambda}^C)\nonumber\\&&+\frac{i}{2}
g\kappa^{EDB}f^{ACE}d^{BDE} A_\mu^C A_\nu^D
{\bar\lambda}^A Y^{\mu\nu} {\bar\lambda}^B
\end{eqnarray}
The two last terms in Eq. (46) help us to make renormalisable NAC $U(N)$ gauge  theory, the first two terms are needed because of gauge transformation rules. These terms are absent from the original lagrangian  because of $\lambda$ redefinition in Ref ~\cite{a1} as we shall  explain in next part. Adding  the above terms to original lagrangian,  the total
lagrangian is given by:
\begin{eqnarray}
L_{total}=L_{original}+L_{Extra}
\end{eqnarray}
Since we add some terms to original lagrangian, we have to modify
the gauge transformation and SUSY transformation. It is easy to show that $L_{total}$ is preserved under following gauge transformation in $U(N)$ group:
\begin{eqnarray}
&&\delta_\phi A_{\mu}^A=-2\partial_{\mu}\phi^A-f^{ABC}\phi^B
A_{\mu}^C,\\&& \delta_\phi
{\bar\lambda}_{\dot\alpha}^A=-f^{ABC}\phi^B
{\bar\lambda}_{\dot\alpha}^C\\&& \delta_\phi
{\lambda}_{\alpha}^A=-f^{ABC}\phi^B-\frac{1}{2}\kappa^{ABC}
d^{ABC}C^{\mu\nu}\sigma_{\nu\alpha\dot\alpha}\pa_\mu\phi^B\bar\lambda^{\dot\alpha
C }\\&& \delta_\phi D^A=-f^{ABC}\phi^B D^C
\end{eqnarray}
Where $\kappa^{ABC}$ is considered  as arbitrary coupling which depends on $A, B, C$ values. The gauge transformation is not canonical because the
transformation of $\lambda$ depends on the non- anti- commutative
parameter $C$. However; the SUSY transformation does not change except
$$\delta\lambda_\alpha^A=i\epsilon_\alpha
D^A+{(\sigma^{\mu\nu}\epsilon)}_\alpha[F_{\mu\nu}^A+\frac{1}{2}iC_{\mu\nu}(\gamma^{ABC}+\frac{1}{2}\kappa^{ABC})d^{ABC}{\bar\lambda}^B{\bar\lambda}^C]$$
So yet we can refer to eqs. (21-23).
In our work in order to renormalise the  NAC  $SU(N)\times U(1)$ gauge theory, we choose
\begin{eqnarray}
\kappa^{ABC}=\xi\gamma^{BAC}c^Ac^Bd^C
\end{eqnarray}
here $\xi$ is  considered as a coefficient.  Then the extra lagrangian is given by:
\begin{eqnarray}
L_{extra}
&=&
\frac{i}{16}\kappa_1 d^{abc}C^{\mu\nu}
( \partial_{\mu} A_\nu^a- \partial_{\nu} A_\mu^a )
\bar\lambda^b \bar\lambda^{c}
- \frac{i}{8}\kappa_1 g  f^{cde} d^{abe} C^{\mu\nu} A_{\mu}^c A_\nu^d
\bar\lambda^a \bar\lambda^{b}
\nonumber \\
&& \qquad
+ \frac{i}{4}\kappa_1 d^{abc}
(\partial_{\mu} \bar\lambda^b Y^{\mu\nu}  \bar\lambda^{c}
- \bar\lambda^b Y^{\mu\nu} \partial_{\mu} \bar\lambda^{c} )
A_\nu^a
-\frac{i}{4}\kappa_1 g  f^{abe} d^{cde}  A_{\mu}^c A_\nu^d
\bar\lambda^a Y^{\mu\nu} \bar\lambda^{b}\nonumber \\&&
+\frac{i}{16}\kappa_3 d^{ab0}C^{\mu\nu}
( \partial_{\mu} A_\nu^a- \partial_{\nu} A_\mu^a )
\bar\lambda^b \bar\lambda^{0}
+ \frac{i}{4}\kappa_3 d^{ab0}
(\partial_{\mu} \bar\lambda^b Y^{\mu\nu}  \bar\lambda^{0}
- \bar\lambda^b Y^{\mu\nu} \partial_{\mu} \bar\lambda^{0} )
A_\nu^a\nonumber \\
&& \qquad
- \frac{i}{4}\kappa_3 g  f^{cde} d^{0be} C^{\mu\nu} A_{\mu}^c A_\nu^d
\bar\lambda^0 \bar\lambda^{b},
\end{eqnarray}
where
\begin{eqnarray}
\kappa_1=\kappa^{abc}=\xi \gamma^{abc},\ \kappa_3=\kappa^{ab0}=2\xi \gamma^{ab0},\ \kappa^{0ab}=\kappa^{a0b}=0.
\end{eqnarray}

In according to Eq.(46), the $Y$ terms leads to new interactions
hence, we have to consider new vertices in 1PI graphs. It means
we display these interactions that  have  been hidden. So,
we  should  calculate 1PI diagrams considering new vertices or we should modify vertices. Finally, we
find that theory is renormalisable.

In this case the new action is given by:
\begin{eqnarray}
S_{total}&=&\int d^4x\Big[-\frac{1}{4}F^{\mu\nu A
}F_{\mu\nu}^A-i{\bar\lambda}^A{\bar\sigma}^\mu{(D_\mu\lambda)}^A+\frac{1}{2}
D^AD^A-\frac{1}{2}id^{ABC}\gamma^{ABC}C^{\mu\nu}F_{\mu\nu}^A{\bar\lambda}^B{\bar\lambda}^C\nonumber\\&&
+\frac{1}{8}\mid
C\mid^2 d^{ABE}d^{CDE}\gamma^{ABCDE}({\bar\lambda}^A{\bar\lambda}^B)({\bar\lambda}^C{\bar\lambda}^D)+L_{Extra}\Big].
\end{eqnarray}

We have to calculate new (1PI) diagrams contributing to the new
terms(those containing both parameter $C$ and $Y$) which are depicted in Figs.~5 and 6.
The results for the new graphs contributing to the new
interaction terms in Eq.~(53) are the same as the $C$ terms so we
shall not give detailed results. For example in order to
calculate Fig. 5-a, we should only change
the NAC vertex $i\gamma^{AJI}d^{AJI}
C^{\mu\nu}\epsilon^{\dot\gamma\dot\delta}(p_1+p_2)_\nu$ in eq. (31)  to
$-\frac{i}{8}\kappa^{JAI}d^{AJI}
C^{\mu\nu}\epsilon^{\dot\gamma\dot\delta}q_\nu+\frac{i}{4}\kappa^{JAI}d^{AJI}({Y^{\mu\nu}})^{\dot\gamma\dot\delta}(p_1-p_2-2k)_\nu$.

The result for the graphs in Fig.~5-a is given by:
\begin{eqnarray}
\Gamma_{5-a-Extra}^{(1)\mu\dot\alpha\dot\beta}=\frac{i}{4}NL\kappa^{BAC}d^{ABC}d^Ac^Bc^C\epsilon^{\dot\alpha\dot\beta}C^{\mu\nu}q_\nu
\end{eqnarray}
Beside, the total  divergent contribution for new graphs in Fig.~5 is
given by:
\begin{eqnarray}
\Gamma_{1PI - Extra\ graph }^{(1)\mu\dot\alpha\dot\beta}&=&-\frac{i}{2}NL\kappa^{abc} d^{abc}\epsilon^{\dot\alpha\dot\beta}C^{\mu\nu}q_\nu\nonumber\\&&
+iNL\kappa^{abc}Ld^{abc}{(Y^{\mu\nu})}^{\dot\alpha\dot\beta}(p_1-p_2)_\nu
\end{eqnarray}

In order to obtain  divergent contribution for 1PI graphs with both
$C$ and $Y$ parameters, one should add  results of  Eq. (39) and Eq. (56)

\begin{eqnarray}
\Gamma_{1PI-total }^{(1)\mu\dot\alpha\dot\beta}&=&(-\frac{1}{2}\kappa^{abc}+\frac{15}{4}\gamma^{abc}) iNLd^{abc}\epsilon^{\dot\alpha\dot\beta}C^{\mu\nu}q_\nu+8\gamma^{0bc}iNLd^{0bc}\epsilon^{\dot\alpha\dot\beta}C^{\mu\nu}q_\nu\nonumber\\&&-\frac{1}{2}\gamma^{a0c}NLd^{a0c}\epsilon^{\dot\alpha\dot\beta}C^{\mu\nu}q_\nu
-\frac{1}{2}\gamma^{ab0}NLd^{ab0}\epsilon^{\dot\alpha\dot\beta}C^{\mu\nu}q_\nu\nonumber\\&&
+(\kappa^{abc}+\frac{1}{2}\gamma^{abc})iNLd^{abc}{(Y^{\mu\nu})}^{\dot\alpha\dot\beta}(p_1-p_2)_\nu\nonumber\\&&
-\gamma^{a0c}iNLd^{a0c}{(Y^{\mu\nu})}^{\dot\alpha\dot\beta}p_{2\nu}
+\gamma^{ab0}iNLd^{ab0}{(Y^{\mu\nu})}^{\dot\alpha\dot\beta}p_{1\nu}
\end{eqnarray}

There are  new graphs(Fig.~6) for one loop corrections of the four point function.
The total contributions as different $SU(N)\times U(1)$ parts corresponds to Fig.~ 5 is given by:
\begin{eqnarray}
\Gamma_{1PI-Extra\ graphs}^{\mu\nu\dot\alpha\dot\beta}&=&-\frac{3}{4}NL\kappa^{edb}gd^{abe}f^{cde}\epsilon^{\dot\alpha\dot\beta}C^{\mu\nu}
-\frac{1}{2}NL\kappa^{ed0}gd^{0de}f^{ace}\epsilon^{\dot\alpha\dot\beta}C^{\mu\nu}\nonumber\\&&
+\frac{3}{2}NL\kappa^{ecd}gd^{cde}f^{abe}{(Y^{\mu\nu})}^{\dot\alpha\dot\beta}
\end{eqnarray}

Then, the total four point 1PI  divergent contribution  is
given by:
\begin{eqnarray}
\Gamma_{1PI-total}^{\mu\nu\dot\alpha\dot\beta}&=&(3\kappa_1+\frac{11}{4}\gamma^{eab})NLgd^{abe}f^{cde}\epsilon^{\dot\alpha\dot\beta}C^{\mu\nu}\nonumber\\&&
+\gamma^{ea0}NLgd^{a0e}f^{cde}\epsilon^{\dot\alpha\dot\beta}C^{\mu\nu}\nonumber\\&&
-\frac{1}{2}\kappa_3NLgd^{0be}f^{cde}\epsilon^{\dot\alpha\dot\beta}C^{\mu\nu}\nonumber\\&&
+(\frac{3}{2}\kappa_1+\frac{1}{2}\gamma^{ecd})NLgd^{cde}f^{abe}{(Y^{\mu\nu})}^{\dot\alpha\dot\beta}
\end{eqnarray}
Fortunately, new terms does not effect on four point function is containing $\mid C\mid^2 (\bar\lambda\bar\lambda)^2$. In order to compute counter terms we should decompose the lagrangian to the $ SU(N)\times U(1)$ parts because some interaction terms such as term which correspond to $(U(1))^{3}$   receive no quantum corrections. It is given by:
\begin{eqnarray}
L_{total}&=&-\frac{1}{2}id^{abc}(\gamma^{abc}-\frac{1}{8}\kappa_1)
C^{\mu\nu}(\partial_\mu A_\nu^a-\partial_\nu A_\mu^a
){\bar\lambda}^b{\bar\lambda}^c\nonumber\\&&
-\frac{1}{2}id^{000}(\gamma^{000}-0) C^{\mu\nu}(\partial_\mu
A_\nu^0-\partial_\nu A_\mu^0
){\bar\lambda}^0{\bar\lambda}^0\nonumber\\&&
-\frac{1}{2}id^{a0c}(\gamma^{a0c}-0)C^{\mu\nu}(\partial_\mu
A_\nu^a-\partial_\nu A_\mu^a
){\bar\lambda}^0{\bar\lambda}^c\nonumber\\&&
-\frac{1}{2}id^{ab0}(\gamma^{ab0}-\frac{1}{8}\kappa_3)C^{\mu\nu}(\partial_\mu
A_\nu^a-\partial_\nu A_\mu^a
){\bar\lambda}^b{\bar\lambda}^0\nonumber\\&&
-\frac{1}{2}id^{0bc}(\gamma^{0bc}-0) C^{\mu\nu}(\partial_\mu
A_\nu^0-\partial_\nu A_\mu^0
){\bar\lambda}^b{\bar\lambda}^c\nonumber\\&&
+\frac{1}{2}ig(\gamma^{eab}-\frac{1}{4}\kappa_1)d^{abe}f^{cde}C^{\mu\nu}A_\mu^c
A_\nu^d\bar\lambda^a\bar\lambda^b\nonumber\\&&
+\frac{1}{2}ig(\gamma^{e0b}-\frac{1}{2}\kappa_3)d^{0be}f^{cde}C^{\mu\nu}A_\mu^c
A_\nu^d\bar\lambda^0\bar\lambda^b\nonumber\\&&
+\frac{1}{2}ig(\gamma^{ea0}-0)d^{a0e}f^{cde}C^{\mu\nu}A_\mu^c
A_\nu^d\bar\lambda^a\bar\lambda^0\nonumber\\&&
+\frac{1}{8}\gamma_0  \mid
C\mid^2d^{abe}d^{cde}({\bar\lambda}^a{\bar\lambda}^b)({\bar\lambda}^c{\bar\lambda}^d)\nonumber\\&&
+\frac{1}{4N}\gamma_1 \mid
C\mid^2({\bar\lambda}^a{\bar\lambda}^a)({\bar\lambda}^b{\bar\lambda}^b)
+\frac{1}{N}\gamma_2 \mid
C\mid^2({\bar\lambda}^a{\bar\lambda}^a)({\bar\lambda}^0{\bar\lambda}^0)\nonumber\\&&
 +\frac{i}{4}\kappa_1
d^{abc}A_\mu^a(\partial_\nu
{\bar\lambda}^bY^{\mu\nu}{\bar\lambda}^c-
{\bar\lambda}^bY^{\mu\nu}\partial_\nu{\bar\lambda}^c)\nonumber\\&&+\frac{i}{4}\kappa_3
d^{ab0}A_\mu^a(\partial_\nu
{\bar\lambda}^bY^{\mu\nu}{\bar\lambda}^0-
{\bar\lambda}^bY^{\mu\nu}\partial_\nu{\bar\lambda}^0)\nonumber\\&&
-\frac{i}{4}\kappa_1 gf^{abe}d^{cde} A_\mu^c A_\nu^d
{\bar\lambda}^a Y^{\mu\nu}{\bar\lambda}^b
\end{eqnarray}
The $C$ and $Y$-dependent part of the action are renormalisable if one
introduces bare fields and couplings according to Eqs.~(26)
and Eqs. (44, 45) as well as:
\begin{eqnarray}
Y^{\mu\nu}_B&=&Z_Y Y^{\mu\nu}, \kappa^{ABC}_B=Z_{\xi}Z_{\gamma^{BAC}}\kappa^{ABC}
\end{eqnarray}
Then in order to find some counter terms we have to introduce the following identities:
\begin{eqnarray}
&&Z_{\gamma^{abc}}Z_CZ_{A}^{1/2}Z_\lambda=\delta_{01}+1,\ \ Z_{\kappa_1}Z_CZ_{A}^{1/2}Z_\lambda=\delta_1+1,\nonumber\\&&
Z_{\gamma^{000}}Z_CZ_{A^0}^{1/2}Z_{\lambda^0}=\delta_{02}+1,\nonumber\\&&
Z_{\gamma^{a0c}}Z_CZ_{A}^{1/2}Z_\lambda^{1/2}Z_{\lambda^0}^{1/2}=\delta_{03}+1,\nonumber\\&&
Z_{\gamma^{ab0}}Z_CZ_{A}^{1/2}Z_\lambda^{1/2}Z_{\lambda^0}^{1/2}=\delta_{04}+1, \ \ Z_{\kappa_3}Z_CZ_{A}^{1/2}Z_\lambda^{1/2}Z_{\lambda^0}^{1/2}=\delta_4+1,\nonumber\\&&
Z_{\gamma^{0bc}}Z_CZ_{A^0}^{1/2}Z_\lambda=\delta_{05}+1, \nonumber\\&&
{Z_g}Z_{\gamma^{abc}}Z_CZ_{A}Z_\lambda=\delta_{06}+1, \ \
{Z_g}Z_{\kappa_1}Z_CZ_{A}Z_\lambda=\delta_6+1,\nonumber\\&&
{Z_g}Z_{\gamma^{e0b}}Z_CZ_{A}Z_{\lambda^0}^{1/2}Z_\lambda^{1/2}=\delta_{07}+1, \ \
{Z_g}Z_{\kappa_3}Z_CZ_{A}Z_{\lambda^0}^{1/2}Z_\lambda^{1/2}=\delta_7+1,\nonumber\\&&
{Z_g}Z_{\gamma^{ea0}}Z_CZ_{A}Z_\lambda^{1/2}Z_{\lambda^0}^{1/2}=\delta_{08}+1,\nonumber\\&&
Z_{\gamma^{0}}Z_{\mid C\mid^2}Z_\lambda^2=\delta_{09}+1,\ \ Z_{\gamma^{1}}Z_{\mid C\mid^2}Z_\lambda^2=\delta_{010}+1, \nonumber\\&&
Z_{\gamma^{2}}Z_{\mid C\mid^2}Z_\lambda Z_{\lambda^0}=\delta_{011}+1,\nonumber\\&&
Z_{\kappa_1}Z_YZ_{A}^{1/2}Z_\lambda=\delta_{12}+1,\
Z_{\kappa_3}Z_YZ_{A}^{1/2}Z_\lambda^{1/2}Z_{\lambda^0}^{1/2}=\delta_{13}+1,\nonumber\\&&
{Z_g}Z_{\kappa_1}Z_YZ_{A}Z_\lambda=\delta_{14}+1,
\end{eqnarray}
Adding the following  counter-term terms to the part of the total
action  and comparing the  expression with the bare action, the
theory should be renormalisable. The full $C$ dependent part of the action can be written as:
\begin{eqnarray}
S&=&\int d^4x \Big[-\frac{1}{2}id^{abc}(\gamma^{abc}(1+\delta_{01})-\frac{1}{8}\kappa_1(1+\delta_{1}))
C^{\mu\nu}(\partial_\mu A_\nu^a-\partial_\nu A_\mu^a
){\bar\lambda}^b{\bar\lambda}^c\nonumber\\&&
-\frac{1}{2}id^{000}(\gamma^{000}(1+\delta_{02})-0) C^{\mu\nu}(\partial_\mu
A_\nu^0-\partial_\nu A_\mu^0
){\bar\lambda}^0{\bar\lambda}^0\nonumber\\&&
-\frac{1}{2}id^{a0c}(\gamma^{a0c}(1+\delta_{03})-0)C^{\mu\nu}(\partial_\mu
A_\nu^a-\partial_\nu A_\mu^a
){\bar\lambda}^0{\bar\lambda}^c\nonumber\\&&
-\frac{1}{2}id^{ab0}(\gamma^{ab0}(1+\delta_{04})-\frac{1}{8}\kappa_3(1+\delta_{4}))C^{\mu\nu}(\partial_\mu
A_\nu^a-\partial_\nu A_\mu^a
){\bar\lambda}^b{\bar\lambda}^0\nonumber\\&&
-\frac{1}{2}id^{0bc}(\gamma^{0bc}(1+\delta_{05})-0) C^{\mu\nu}(\partial_\mu
A_\nu^0-\partial_\nu A_\mu^0
){\bar\lambda}^b{\bar\lambda}^c\nonumber\\&&
+\frac{1}{2}ig(\gamma^{eab}(1+\delta_{06})-\frac{1}{4}\kappa_1(1+\delta_{6}))d^{abe}f^{cde}C^{\mu\nu}A_\mu^c
A_\nu^d\bar\lambda^a\bar\lambda^b\nonumber\\&&
+\frac{1}{2}ig(\gamma^{e0b}(1+\delta_{07})-\frac{1}{2}\kappa_3(1+\delta_{7}))d^{0be}f^{cde}C^{\mu\nu}A_\mu^c
A_\nu^d\bar\lambda^0\bar\lambda^b\nonumber\\&&
+\frac{1}{2}ig(\gamma^{ea0}(1+\delta_{08})-0)d^{a0e}f^{cde}C^{\mu\nu}A_\mu^c
A_\nu^d\bar\lambda^a\bar\lambda^0\nonumber\\&&
+\frac{1}{8}(1+\delta_{09})\gamma_0  \mid
C\mid^2d^{abe}d^{cde}({\bar\lambda}^a{\bar\lambda}^b)({\bar\lambda}^c{\bar\lambda}^d)\nonumber\\&&
+\frac{1}{4N}(1+\delta_{010})\gamma_1 \mid
C\mid^2({\bar\lambda}^a{\bar\lambda}^a)({\bar\lambda}^b{\bar\lambda}^b)
+\frac{1}{N}(1+\delta_{011})\gamma_2 \mid
C\mid^2({\bar\lambda}^a{\bar\lambda}^a)({\bar\lambda}^0{\bar\lambda}^0)\nonumber\\&&
 +\frac{i}{4}(1+\delta_{12})\kappa_1
d^{abc}A_\mu^a(\partial_\nu
{\bar\lambda}^bY^{\mu\nu}{\bar\lambda}^c-
{\bar\lambda}^bY^{\mu\nu}\partial_\nu{\bar\lambda}^c)\nonumber\\&&+\frac{i}{4}(1+\delta_{13})\kappa_3
d^{ab0}A_\mu^a(\partial_\nu
{\bar\lambda}^bY^{\mu\nu}{\bar\lambda}^0-
{\bar\lambda}^bY^{\mu\nu}\partial_\nu{\bar\lambda}^0)\nonumber\\&&
-\frac{i}{4}(1+\delta_{14})\kappa_1 gf^{abe}d^{cde} A_\mu^c A_\nu^d
{\bar\lambda}^a Y^{\mu\nu}{\bar\lambda}^b\Big].
\end{eqnarray}

where in order to renormalise the $C$-dependent part of the
action we have to obtain  the values of $\delta_i$ by solving the following equations:
\begin{eqnarray}
&&\Gamma_{1PI-total}^{(1)\mu\dot\alpha\dot\beta}+\Gamma_{C.T}^{\mu\dot\alpha\dot\beta}=0,\\&&
\Gamma_{1PI-total}^{(1)\mu\nu\dot\alpha\dot\beta}+\Gamma_{C.T}^{\mu\nu\dot\alpha\dot\beta}=0,\\&&
\Gamma_{1PI-total}^{(1)\dot\alpha\dot\beta\dot\delta\dot\gamma}+\Gamma_{C.T}^{\dot\alpha\dot\beta\dot\delta\dot\gamma}=0,
\end{eqnarray}
Where $\Gamma_{C.T}$s come from  counter terms (Appendix E).
Then, using $Z_g,\ Z_A,\ Z_\lambda$ for $SU(N)$ sector, and $Z_{g_0},\ Z_{A^0},\ Z_{\lambda^0}$ for $U(1)$ sector, we obtain the renormalisation constants as :
\begin{eqnarray}
&& Z_{\xi}^{(1)}=-\frac{2}{\xi}NL, \ Z_{h}^{(1)}=-NL
\end{eqnarray}

As it is understood because $\xi$ is renormalised, so  $\kappa^{ABC}$ should be renormalised.
Moreover  we  obtain
\begin{eqnarray}
&&Z_C=Z_Y= Z_{\mid
C\mid^2}=1 \\&&
(Z_C)^2=Z_{\mid C\mid^2}=1
\end{eqnarray}
It is found that our result is compatible with Ref~\cite{a11,a13}. Of course a natural expectation would be that $Z_C=Z_Y$ , because
of Eq. (36) we know that $Y\propto C$.

We demonstrate that the theory is renormalisable and the N = 1/2 supersymmetric  as well as  NAC $SU(N)\times U(1)$ pure  gauge theory is preserved.
This point also has been concluded  in Ref~\cite{a11} and suggested in Ref~\cite{a6}, where was supposed to be correct to all orders. We also arrive at the conclusion that it is not needed to renormalise the non-anticommutativity parameter $C$. Beside our full lagrangian is the same form as derived from non-anticommutative superspace, however  $Z_{\kappa^{ABC}}$ and $Z_{\gamma^{ABC}}$ depends on whether $A,B,C$ are $SU( N )$ indices or $U(1)$ indices.
It seems to imply that the renormalised theory is not $U( N )$ non-anticommutative theory any more.
Because the $U( N )$  structure is broken by renormalisation.

In order to clarify $L_{Extra}$ we  redefine the component
$\lambda$ as
\begin{eqnarray}
\lambda^A\longrightarrow\lambda^A-\frac{1}{4}\kappa^{ABC}
d^{ABC}C^{\mu\nu} A_\mu^B \sigma_{\nu} \bar\lambda^{C}
\end{eqnarray}

Then, put it in Eq.~(18), and  obtain Eq.~(55)(in other words $L_{total}$ is result of $\lambda$ redefinition in $L_{original}$).The $\lambda$ redefinition just affects the gaugino kinetic term. Our redefinition is opposite to Refs~\cite{a1,a20}. They have redefined $\lambda$ in order to make gauge transformations be canonical; however it causes theory unrenormalisable because  in that case  some terms are been hidden in the lagrangian. In order to reverse process we should add hidden terms by hand or come back to original definition of  $\lambda$  Eq.~(71); however we  lose  gauge canonical transformation. Beside, because $\kappa^{ABC}$ is   obtained  renormalised, the redefinition  of $\lambda^A$ is called $\hat\lambda^A$ should be renormalised. Finally divergent field redefinition in Ref~\cite{a11} is reinterpreted as renormalised $\hat\lambda^A$. Our results show it is not needed to deform the classical action if  one do not  use the field redefinition of Ref~\cite{a1}.

\subsection{ discussion on   $\xi\longrightarrow 0$}
It is worthwhile to investigate  our results in the case of limit $\xi\longrightarrow 0$. Indeed in this case, 1PI graphs from new terms are vanished. We would like to present this claim as follow. Instance we know that
$ \Gamma_{1-Extra}^{(1)\mu\dot\alpha\dot\beta}(\xi)$ comes from extra lagrangian. Then  total divergent contribution for three point function would be finite if  we  add some counter terms as follow:

\begin{eqnarray}
\Gamma_{1PI-total}^{(1)\mu\dot\alpha\dot\beta}=\Gamma_{1-original}^{(1)\mu\dot\alpha\dot\beta}+ \Gamma_{1-Extra}^{(1)\mu\dot\alpha\dot\beta}(\xi)+\Gamma_{C.T}^{\mu\dot\alpha\dot\beta}=0
\end{eqnarray}
we expect to obtain $Z_C=Z_Y=Z_{\mid C\mid^2}=1$ (it means  the NAC structure would be preserved under the procedure of renormalisation). Then, the above equation leads to:
\begin{eqnarray}
&&
\delta_{1}=-(\frac{2}{\xi}+4)NL,\\&&
\delta_{02}=0,\\&&
\delta_{03}=\delta_{04}=0,\\&&
\delta_4=-\frac{2}{\xi}NL,\\&&
\delta_{05}=-8NL\\&&
\delta_{9}=-(\frac{2}{\xi}+4)NL,\\&&
\delta_{11}=-\frac{2}{\xi}NL
\end{eqnarray}
 Note that second terms in eqs. (73) and (78) are related to divergent contribution from new 1PI graphs.
In order to find  $Z_\xi$ we have to benefit from eqs. (63). Hence $Z_\xi$ up to one order is given by:
\begin{eqnarray}
Z_{\xi}=(1-\frac{2}{\xi}NL-4NL)(1+4NL)=1-\frac{2}{\xi}NL,
\end{eqnarray}
Now, in the case of $\xi\longrightarrow 0$  eqs. (73-79) result to
\begin{eqnarray}
&&
\delta_{1}=-\frac{2}{\xi}NL,\
\delta_{02}=0,\
\delta_{03}=\delta_{04}=0,\\&&
\delta_4=-\frac{2}{\xi}NL,\
\delta_{05}=-8NL,\
\delta_{9}=-\frac{2}{\xi}NL,\
\delta_{11}=-\frac{2}{\xi}NL.
\end{eqnarray}
 In fact we have neglected the divergent contributions from new graphs, however; $Z_\xi$ has not been changed. This event could be checked for four point function as well. In this case we obtain the results of Ref~\cite{a11,a10}, but because  $Z_\xi\neq0$ it easy to show it is not necessary to define a nonlinear divergence field redefinition.
If we consider
\begin{eqnarray}
\hat\lambda^A=\lambda^A-\frac{1}{4}\kappa^{ABC}
d^{ABC}C^{\mu\nu} A_\mu^B \sigma_{\nu} \bar\lambda^{C}
\end{eqnarray}
Then the variation of $\lambda^A$ is written by
\begin{eqnarray}
\delta\lambda^A=-\frac{1}{4}\kappa^{ABC}
d^{ABC}C^{\mu\nu} A_\mu^B \sigma_{\nu} \bar\lambda^{C}
\end{eqnarray}
Consequently, after the procedure of renormalization, the nonlinear divergent field redefinition in Ref~\cite{a11} automatically is generated:
\begin{eqnarray}
\delta\lambda^A=\frac{1}{2}NL\gamma^{BAC}c^Ac^Bd^C
d^{ABC}C^{\mu\nu} A_\mu^B \sigma_{\nu} \bar\lambda^{C}
\end{eqnarray}
\section{Conclusion}
We have compute  1pI corrections for the pure $N=\frac{1}{2}$
supersymmetric $SU(N)\times U(1) $ gauge theory at one loop order. We have
proved the theory is renormalisable up one loop order using a
standard way of  renormalisation if one adds some new terms to the
original lagrangian. We have shown it is possible to interrupt
these are hidden terms because of $\lambda$ redefinition. it is
worth to investigate if it is possible to show  that the problems
which arise in renormalisation of $N=\frac{1}{2}$ supersymmetric
theories comes from the redefind vector  superfield.

We have shown there is not need to define divergent redefinition of $\lambda$. Moreover we suggest  all
works which have been done based  on divergent field redefinition
should be reviewed. We have used the $N=\frac{1}{2}$ $U(N)$
gauge group action because as discussed in \cite{a21}, just non-anticommutative theory with $U(N)$ gauge group is  well-defined. Moreover it is worth to investigate the renormalization of theory at higher loops or including chiral
matter in the standard form of renormalisation method. We guess the problem of renormalisation  of non-anticommutative theory  at component formalism  is because of  $\lambda$ redefinition in ~\cite{a1}.
\section{Acknowledgements}
It is a pleasure to thank F. Yagi  for helpful discussions and N. Ghasempour for checking calculations.
\appendix
\section{: New Algebra For Non Commutative Parameters C And Y}
We have found the new properties for C and Y parameters that we
have made frequent use in our calculations. $C^{\mu\nu}$ is
related to the non-anti-commutativity parameter $C^{\alpha\beta}$
by:
\begin{eqnarray}
C^{\mu\nu}&=&C^{\alpha\beta}\epsilon_{\beta\gamma}\sigma_\alpha^{\mu\nu\
\gamma}
\end{eqnarray}
also, we have:
\begin{eqnarray}
C^{\alpha\beta}&=&\frac{1}{2}\epsilon^{\alpha\gamma}\sigma_\gamma^{\mu\nu\
\beta} C_{\mu\nu},
\end{eqnarray}
where
\begin{eqnarray}
&&C^{\mu\nu}\sigma_{\nu\alpha\dot\beta}=\epsilon_{\alpha\beta}C^{\beta\gamma}\sigma^\mu_{\gamma\dot\beta}
~,\\&&
C^{\mu\nu}{\bar\sigma}_\nu^{\dot\alpha\beta}=C^{\beta\alpha}\epsilon_{\alpha\gamma}{\bar\sigma}^{\mu\dot\alpha\gamma}
\end{eqnarray}
we have used the following notations:
\begin{eqnarray}
&&C^{\alpha\gamma}\epsilon_{\gamma\beta}=C^\alpha ~_\beta ~,\\&&
\epsilon_{\beta\gamma}C^{\gamma\alpha}=C_\beta~ ^\alpha ~,\\&&
C^\alpha ~_\beta =-{(C_\beta~ ^\alpha)}^T
\end{eqnarray}
where in last equation the symbol of T is used for  transposed.
Also for Y parameter we have:
\begin{eqnarray}
&&{(Y^{\mu\nu})}^{\dot\alpha}_{\dot\theta}=C^{\mu\rho}g_{\rho\lambda}({\bar\sigma}^{\lambda\nu})^{\dot\alpha}_{\dot\theta}~,\\&&
{(Y^{\mu\nu})}^{\dot\alpha}_{\dot\theta}{(Y_{\mu\nu})}^{\dot\beta}_{\dot\gamma}
=\frac{1}{4}[\delta^{\dot\alpha}_{\dot\theta}\delta^{\dot\beta}_{\dot\gamma}-2\delta^{\dot\alpha}_{\dot\gamma}\delta^{\dot\beta}_{\dot\theta}]\mid
C\mid^2 ~,\\&& Tr[~\mid Y\mid^2~]=-\frac{3}{2}\mid C\mid^2~,\\&&
{(Y^{\mu\nu})}^{\dot\alpha}_{\dot\theta}\sigma_{\nu\gamma\dot\gamma}=\frac{1}{2}[-2C^\alpha
~_\theta{\bar\sigma}^{\mu\dot\alpha\theta}\epsilon_{\alpha\gamma}\epsilon_{\dot\theta\dot\gamma}+C_\gamma~
^\theta{\sigma^r}_{\theta\dot\gamma}\delta^{\dot\alpha}_{\dot\theta}]~,\\&&
{(Y^{\mu\nu})}^{\dot\alpha}_{\dot\theta}{\bar\sigma}_\nu
~^{\dot\gamma\gamma}=\frac{1}{2}C^\gamma
~_\theta[-2{\bar\sigma}^{\mu\dot\alpha\theta}\delta^{\dot\gamma}_{\dot\theta}+{\bar\sigma}^{\mu\dot\gamma\theta}\delta^{\dot\alpha}_{\dot\theta}]~,\\&&
{(Y^{\mu\nu})}^{\dot\alpha}_{\dot\theta}{\bar\sigma}_\mu
~^{\dot\gamma\gamma}\sigma_{\nu\delta\dot\delta}=C^\gamma
~_\delta[2\epsilon^{\dot\alpha\dot\gamma}\epsilon_{\dot\theta\dot\delta}+\delta^{\dot\gamma}_{\dot\delta}\delta^{\dot\alpha}_{\dot\theta}]~,\\&&
{(Y^{\mu\nu})}^{\dot\alpha}_{\dot\theta}{({\bar\sigma}_{\nu\mu})}^{\dot\beta}_{\dot\delta}=0
~,\\&& {(Y^{\mu\nu})}^{\dot\alpha}_{\dot\theta}g_{\mu\nu}=0 ~,\\&&
{(Y^{\mu\nu})}^{\dot\alpha}_{\dot\alpha}=0,~ (Y\ is\ traceless).
\end{eqnarray}
\section{: Feynman Parameters}
\begin{eqnarray}
\frac{1}{ab}&=&\int_0^1dx\frac{1}{[(1-x)b+xa]^2}\\
\frac{1}{a^nb}&=&n\int_0^1dx\frac{x^{n-1}}{[(1-x)b+xa]^{n+1}}\\
\frac{1}{abc}&=&2\int_0^1dx\int_0^1dy\frac{x}{[axy+bx(1-y)+c(1-x)]^3}\nonumber\\
&=&2\int_0^1du\int_0^{1-u}dw\frac{1}{[aw+b(1-u-w)+cu]^3}\\
\frac{1}{a_1a_2...\ a_n}&=&(n-1)!\int_0^1dx_ndx_{n-1}... \
dx_2\nonumber\\&& \times\frac{x_n^{n-2}x_{n-1}^{n-3}...\
x_3^1x_2^0}{[(1-x_n)a_n
+x_n[(1-x_{n-1})a_{n-1}+x_{n-1}[...+x_3[(1-x_2)a_2+x_2a_1]]...]^n}\nonumber\\
\end{eqnarray}
\section{: The d Dimensional Integrals In Minkowski Space}
\begin{eqnarray}
&&\int\frac{d^dl}{(2\pi)^d}\frac{1}{[l^2-\Delta]^n}=\frac{(-1)^ni}{(4\pi)^{\frac{d}{2}}}\frac{\Gamma(n-\frac{d}{2})}{\Gamma(n)}(\frac{1}{\Delta})^{n-\frac{d}{2}}\\&&
\int\frac{d^dl}{(2\pi)^d}\frac{l^2}{[l^2-\Delta]^n}=\frac{(-1)^{n-1}i}{(4\pi)^{\frac{d}{2}}}\frac{d}{2}\frac{\Gamma(n-\frac{d}{2}-1)}{\Gamma(n)}(\frac{1}{\Delta})^{n-\frac{d}{2}-1}\\&&
\int\frac{d^dl}{(2\pi)^d}\frac{l^\mu
l^\nu}{[l^2-\Delta]^n}=\frac{(-1)^{n-1}i}{(4\pi)^{\frac{d}{2}}}\frac{g^{\mu\nu}}{2}\frac{\Gamma(n-\frac{d}{2}-1)}{\Gamma(n)}(\frac{1}{\Delta})^{n-\frac{d}{2}-1}\\&&
\int\frac{d^dl}{(2\pi)^d}\frac{(l^2)^2}{[l^2-\Delta]^n}=\frac{(-1)^ni}{(4\pi)^{\frac{d}{2}}}\frac{d(d+2)}{4}\frac{\Gamma(n-\frac{d}{2}-1)}{\Gamma(n)}(\frac{1}{\Delta})^{n-\frac{d}{2}-2}\\&&
\int\frac{d^dl}{(2\pi)^d}\frac{l^\mu l^\nu l^\rho
l^\sigma}{[l^2-\Delta]^n}=\frac{(-1)^ni}{(4\pi)^{\frac{d}{2}}}\frac{\Gamma(n-\frac{d}{2}-2)}{\Gamma(n)}(\frac{1}{\Delta})^{n-\frac{d}{2}-2}\nonumber\\&&
\times\frac{1}{4}(g^{\mu\nu}g^{\rho\sigma}+g^{\mu\rho}g^{\nu\sigma}+g^{\mu\sigma}g^{\nu\rho}).
\end{eqnarray}
\section{: Feynman Rules}
Here, we collect Feynman rules would be used in order to calculate 1PI digrams for current theory.
propagators for each field could be as follow:

Gauge field $A_\mu^A$: $$-\frac{ig_{\mu\nu}}{p^2},$$

Gaugino field $\lambda_{\alpha}^{A}$: $$\frac{ip_\mu\sigma_{\alpha\dot\alpha}^\mu}{p^2},$$

Auxiliary Boson field $D^A$: this field could not be propagated.

Scalar field $F^A$: this field could not be propagated.

Spinor field $\psi_{\alpha}^A$: $$\frac{ip_\mu\sigma_{\alpha\dot\alpha}^\mu}{p^2},$$

Scalar field $\phi^A$: $$\frac{i}{p^2}$$

Vertices comes from each interaction in the theory so, we have some vertices as follow:

three- Gauge coupling $A_\mu^a, A_\nu^b, A_\rho^c$ with momentum $k, p, q$ respectively:
$$gf^{abc}[g^{\mu\nu}(k-p)^\rho+g^{\nu\rho}(p-q)^\mu+g^{\rho\mu}(q-k)^\nu],$$

four- Gauge coupling $A_\mu^a, A_\nu^b, A_\rho^c, A_\sigma^d$:
$$-ig^2[f^{abe}f^{cde}(g^{\mu\rho}g^{\nu\sigma}-g^{\mu\sigma}g^{\nu\rho})$$$$+f^{ace}f^{bde}(g^{\mu\nu}g^{\rho\sigma}-g^{\mu\sigma}g^{\nu\rho})$$$$+f^{ade}f^{bce}(g^{\mu\nu}g^{\rho\sigma}-g^{\mu\rho}g^{\nu\sigma})]$$

Gaugion $\bar\lambda_{\dot\alpha}^A$ -Gaugino $\lambda_{\alpha}^C$ -Gauge $A_\mu^B$ vertex: $$-gf^{ABC}\bar\sigma^{\mu\dot\alpha\alpha}$$

NAC in Gauge $A_\mu^A$ with momentum $k_\nu$ -Gaugino $\bar\lambda_{\dot\alpha}^B$ -Gaugino $\bar\lambda_{\dot\beta}^C$ vertex:
$$id^{ABC}\gamma^{ABC}C^{\mu\nu}\epsilon^{\dot\alpha\dot\beta}k_\nu$$

NAC in Gauge $A_\mu^C$ -Gauge $A_\nu^D$  -Gaugino $\bar\lambda_{\dot\alpha}^A$ -Gaugino $\bar\lambda_{\dot\beta}^B$ vertex:
$$-\frac{1}{2}gd^{ABE}\gamma^{EAB}f^{CDE}C^{\mu\nu}\epsilon^{\dot\alpha\dot\beta}$$

NAC in four Gauginos $(\bar\lambda_{\dot\alpha}^A\bar\lambda_{\dot\beta}^B)(\bar\lambda_{\dot\gamma}^C\bar\lambda_{\dot\delta}^D)$ vertex:$$\frac{i}{8}\mid C\mid^2d^{ABE}d^{CDE}\gamma^{ABCDE}\epsilon^{\dot\alpha\dot\beta}\epsilon^{\dot\gamma\dot\delta}$$

For total lagrangian we can write
\begin{eqnarray}
L_{total}=L_{C=0}+L_{C}+L_{Extra}
\end{eqnarray}
So, Feynman rule would be:
\begin{itemize}
\item $A_\mu^A$ gauge, $\bar\lambda_{\dot\alpha}^B$ gaugino, $\bar\lambda_{\dot\beta}^C$ gaugino in NAC parameter $C$:
\begin{eqnarray}
&&iC^{\mu\nu}q_\nu\epsilon^{\dot\alpha\dot\beta}d^{ABC}\gamma^{ABC}
-\frac{i}{8}XC^{\mu\nu}q_\nu\epsilon^{\dot\alpha\dot\beta}d^{ABC}\kappa^{BAC}\nonumber\\&&+\frac{i}{4}{Y^{\mu\nu}}^{\dot\alpha\dot\beta}(p_1-p_2)_\nu d^{ABC}\kappa^{BAC}
\end{eqnarray}
\item $A_\mu^C$ gauge, $A_\mu^D$ gauge, $\bar\lambda_{\dot\alpha}^A$ gaugino, $\bar\lambda_{\dot\beta}^B$ gaugino in NAC parameter $C$:
\begin{eqnarray}
&&\frac{1}{2} gC^{\mu\nu}\epsilon^{\dot\alpha\dot\beta}d^{ABE}f^{CDE}\gamma^{EAB}
-\frac{1}{4}gC^{\mu\nu}\epsilon^{\dot\alpha\dot\beta}d^{BDE}f^{ACE}\kappa^{EDB}\nonumber\\&&+\frac{1}{2}{Y^{\mu\nu}}^{\dot\alpha\dot\beta} d^{BDE}f^{ACE}\kappa^{EDB}
\end{eqnarray}
\end{itemize}
\section{: Counter Terms}
We have made use from the follow Counter Terms in the procedure of the renormalization:
\begin{eqnarray}
\Gamma_{C.T}^{\mu\dot\alpha\dot\beta}&=&
+id^{abc}(\delta_{01}\gamma^{abc}-\frac{\delta_{1}}{8}\kappa_1)
\epsilon^{\dot\alpha\dot\beta}C^{\mu\nu}q_\nu\nonumber\\&&
+id^{000}(\delta_{02}\gamma^{000}-0) \epsilon^{\dot\alpha\dot\beta}C^{\mu\nu}q_\nu\nonumber\\&&
+id^{a0c}(\delta_{03}\gamma^{a0c}-0)\epsilon^{\dot\alpha\dot\beta}C^{\mu\nu}q_\nu\nonumber\\&&
+id^{ab0}(\delta_{04}\gamma^{ab0}-\frac{\delta_{4}}{8}\kappa_3)\epsilon^{\dot\alpha\dot\beta}C^{\mu\nu}q_\nu\nonumber\\&&
+id^{0bc}(\delta_{05}\gamma^{0bc}-0) \epsilon^{\dot\alpha\dot\beta}C^{\mu\nu}q_\nu\nonumber\\&&
+\frac{i}{4}\delta_{12}
d^{abc}\kappa_1{(Y^{\mu\nu})}^{\dot\alpha\dot\beta}(p_1-p_2)_\nu\nonumber\\&&
+\frac{i}{4}\delta_{13}
d^{ab0}\kappa_3{(Y^{\mu\nu})}^{\dot\alpha\dot\beta}(p_1-p_2)_\nu\\ \
\Gamma_{C.T}^{\mu\nu\dot\alpha\dot\beta}&=&\frac{1}{2}(\delta_{06}\gamma^{eab}-\frac{\delta_{6}}{4}\kappa_1)d^{abe}f^{cde}\epsilon^{\dot\alpha\dot\beta}C^{\mu\nu}\nonumber\\&&
+\frac{1}{2}(\delta_{07}\gamma^{e0b}-\frac{\delta_{7}}{2}\kappa_3)d^{0be}f^{cde}\epsilon^{\dot\alpha\dot\beta}C^{\mu\nu}\nonumber\\&&
+\frac{1}{2}(\delta_{08}\gamma^{ea0}-0)d^{a0e}f^{cde}\epsilon^{\dot\alpha\dot\beta}C^{\mu\nu}\nonumber\\&&
+\frac{1}{4}\delta_{14} gf^{abe}d^{cde}\kappa_1
{(Y^{\mu\nu})}^{\dot\alpha\dot\beta}\\
\Gamma_{C.T}^{\dot\alpha\dot\beta\dot\delta\dot\gamma}&=&\frac{\delta_{09}}{8}i\gamma_0\epsilon^{\dot\alpha\dot\beta}\epsilon^{\dot\delta\dot\gamma}d^{abe}d^{cde}
\mid  C\mid^2+\frac{\delta_{010}}{4N}i\gamma_1\epsilon^{\dot\alpha\dot\beta}\epsilon^{\dot\delta\dot\gamma}
\mid C\mid^2\nonumber\\&&
+\frac{\delta_{011}}{N}i\gamma_2\epsilon^{\dot\alpha\dot\beta}\epsilon^{\dot\delta\dot\gamma}\mid C\mid^2
\end{eqnarray}

\section{: U(N) Group Identities}
\begin{eqnarray}
&&f^{ABC}=-f^{ACB}=...,\  d^{ABC}=d^{ACB}=...\nonumber
\\&&
f^{CAD}f^{DBC}=-Nc^A\delta^{AB},\nonumber\\&&
d^{CAD}d^{DBC}=Nd^A\delta^{AB},\nonumber\\&&
f^{CAD}d^{DBC}=0,\nonumber\\&&
f^{DAE}f^{EBF}f^{FCD}=-\frac{N}{2}f^{ABC},\nonumber\\&&
d^{DAE}f^{EBF}f^{FCD}=-\frac{N}{2}d^{ABC}d^A c^B c^C,\nonumber\\&&
f^{DAE}d^{EBF}d^{FCD}=\frac{N}{2}f^{ABC},\nonumber\\&&
d^{IAJ}f^{JBK}f^{KCL}f^{LDI}=-\frac{N}{4}[d^{ABE}f^{CDE}+f^{ABE}d^{CDE}]d^A c^B c^C c^D,\nonumber\\&&
d^{ABCD}=Tr[F^AF^BD^CD^D]=c^Ac^B[\frac{1}{2}c^Cc^D(\delta_{AC}\delta_{BD}-\delta_{AB}\delta_{CD}+\delta_{AD}\delta_{BC})\nonumber\\&&+\frac{N}{8}d^Cd^D(-f^{ABE}f^{CDE}-f^{ACE}f^{BDE}-d^{ABE}d^{CDE}-d^{ACE}d^{BDE})],\nonumber\\&&
\tilde d^{abcd}=Tr[F^aD^c F^b D^d],
\end{eqnarray}
Where $d^A=1+\delta^{0A}$ , $c^A=1-\delta^{0A}$. The gauge index of U(N) runs A = 0, 1, ..., $N^2-1$ where A = 0 corresponds to overall U(1)
while A = a = 1, ..., $N^2-1$ corresponds to SU(N).
For the simplicity of the calculation, we introduce the matrix $F^A$ and $D^A$, whose component
is given by $f^{ABC}$ and $d^{ABC}$ as$$(F^A)_{BC}=f^{BAC},\  (D^A)_{BC}=d^{BAC}$$
Taking into account that $f^{ABC}$  is totally antisymmetric tensor and $d^{ABC}$ is totally symmetric
tensor.
\section{:Sigma Matrices}
\begin{eqnarray}
&&\sigma^\mu_{\alpha\dot\alpha}{\bar\sigma}_\mu^{\dot\beta\beta}=-2\delta_\alpha^\beta\delta^
{\dot\beta}_{\dot\alpha}\nonumber\\&&
\sigma^\mu_{\alpha\dot\alpha}\sigma_{\mu\beta\dot\beta}=-2\epsilon_{\alpha\beta}\epsilon_
{\dot\alpha\dot\beta}\nonumber\\&&
{\bar\sigma}^{\mu\dot\alpha\alpha}{\bar\sigma}_\mu^{\dot\beta\beta}=-2\epsilon^{\alpha\beta}\epsilon^
{\dot\alpha\dot\beta}\nonumber\\&&
(\sigma^\mu{\bar\sigma}_\mu)_\alpha^\beta=-4\delta_\alpha^\beta
\nonumber\\&&
({{\bar\sigma}^\mu\sigma_\mu})^{\dot\alpha}_{\dot\beta}=-4\delta^{\dot\alpha}_{\dot\beta}\nonumber
\\&&
({\sigma^\mu{\bar\sigma}^\nu})_\alpha^\beta=2(\sigma^{\mu\nu})_\alpha^\beta-g^{\mu\nu}\delta_\alpha^\beta
\nonumber\\&&
({\bar\sigma}^\mu\sigma^\nu)^{\dot\alpha}_{\dot\beta}=2({\bar\sigma}^{\mu\nu})^{\dot\alpha}_{\dot\beta}
-g^{\mu\nu}\delta^{\dot\alpha}_{\dot\beta}\nonumber \\&&
(\sigma^{\mu\nu})_\alpha^\beta(\sigma_{\mu\nu})_\rho^\kappa=2\delta_\alpha^\kappa\delta_\rho^\beta-
\delta_\alpha^\beta\delta_\rho^\kappa\nonumber \\&&
({\bar\sigma}^{\mu\nu})^{\dot\alpha}_{\dot\beta}({\bar\sigma}_{\mu\nu})^{\dot\rho}_{\dot\kappa}=
2\delta^{\dot\alpha}_{\dot\kappa}\delta^{\dot\rho}_{\dot\beta}-\delta^{\dot\alpha}_{\dot\beta}
\delta^{\dot\rho}{\dot\kappa}\nonumber \\&&
Tr(\sigma^\mu{\bar\sigma}^\nu)=Tr({\bar\sigma}^\mu\sigma^\nu)=-2g^{\mu\nu}\nonumber\\&&
Tr(\sigma^{\mu\nu})=Tr({\bar\sigma}^{\mu\nu})=0.
\end{eqnarray}

\begin{figure}[ht]
\centerline{\includegraphics[width=13cm]{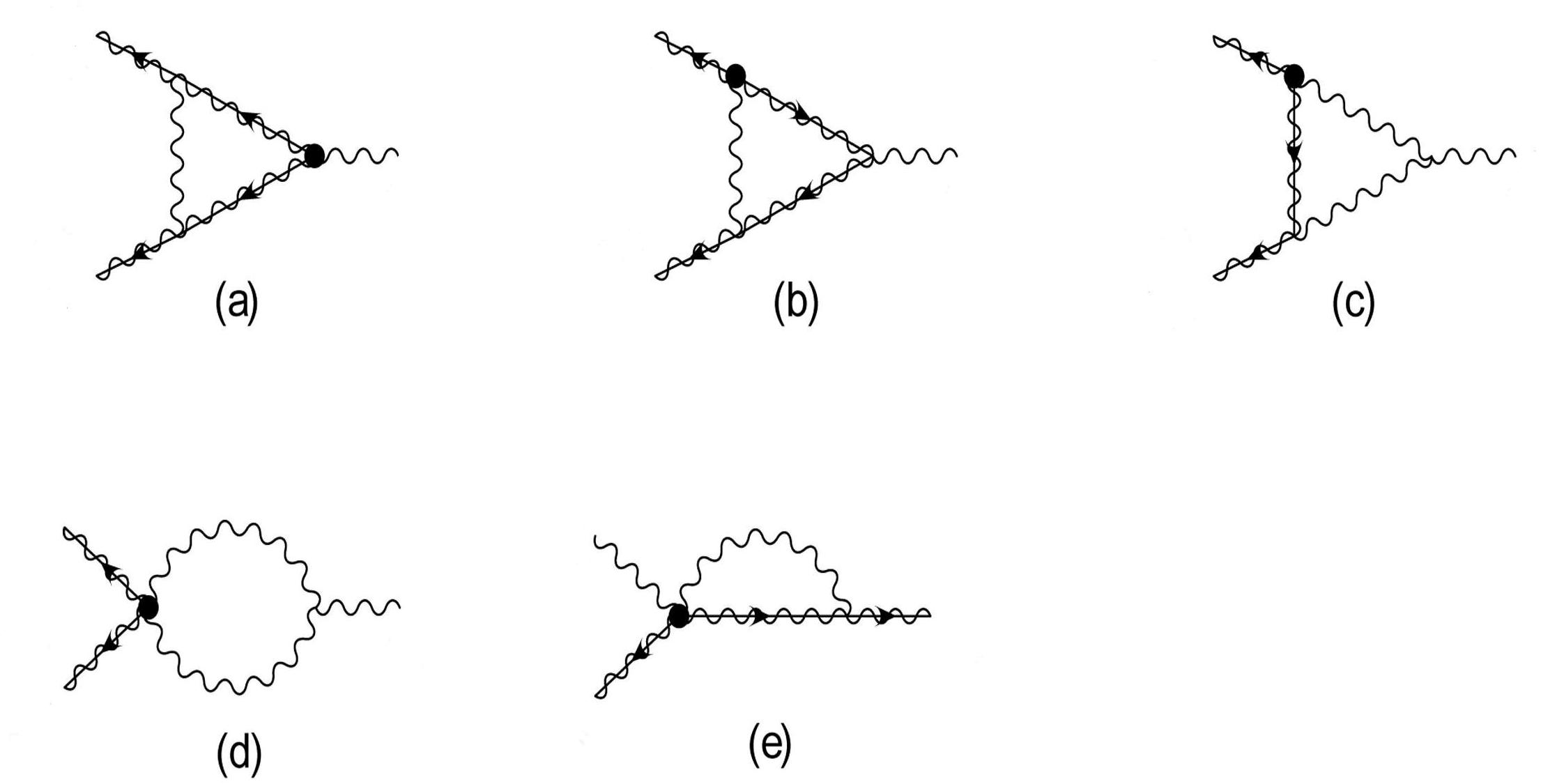}}
\caption{\label{fig1}\small 1PI three point function diagrams  with one gauge, two gaugino lines; the black circle represents the positions of  NAC parameter $C$.}
\end{figure}
\begin{figure}[ht]
\centerline{\includegraphics[width=15cm]{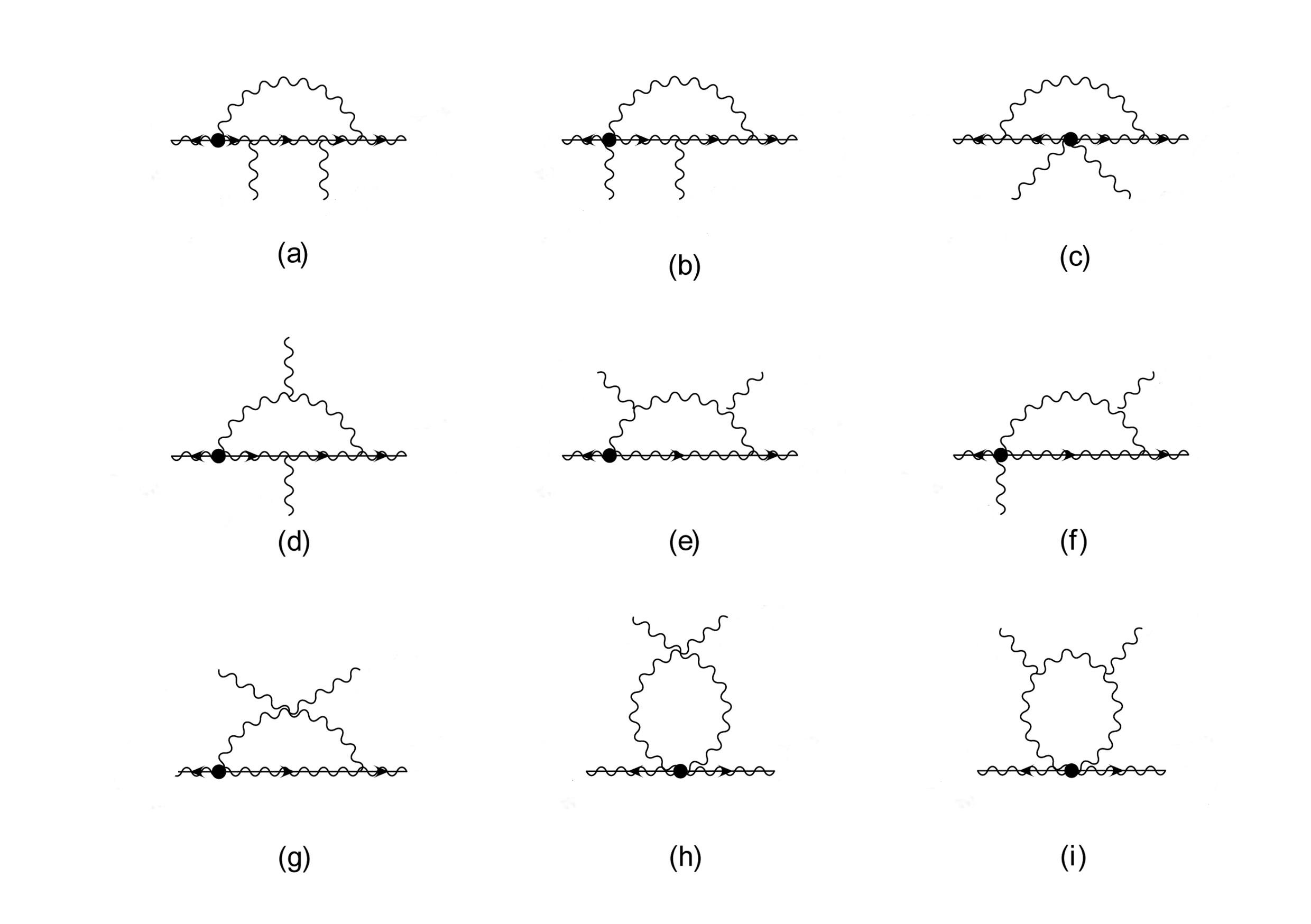}}
\caption{\label{fig1}\small 1PI  four point  function   diagrams with two gauge and two gaugino lines; the black circle represents the positions of a  NAC parameter $C$.}
\end{figure}
\begin{figure}[ht]
\centerline{\includegraphics[width=10cm]{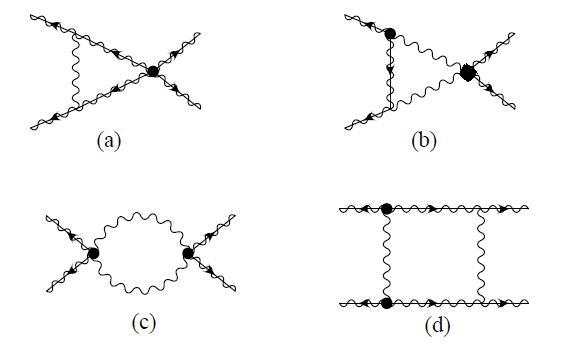}}
\caption{\label{fig1}\small 1PI  four point  function diagrams with four  gaugino lines; the black circle represents the positions of a NAC $C$.}
\end{figure}
\begin{figure}[ht]
\centerline{\includegraphics[width=10cm]{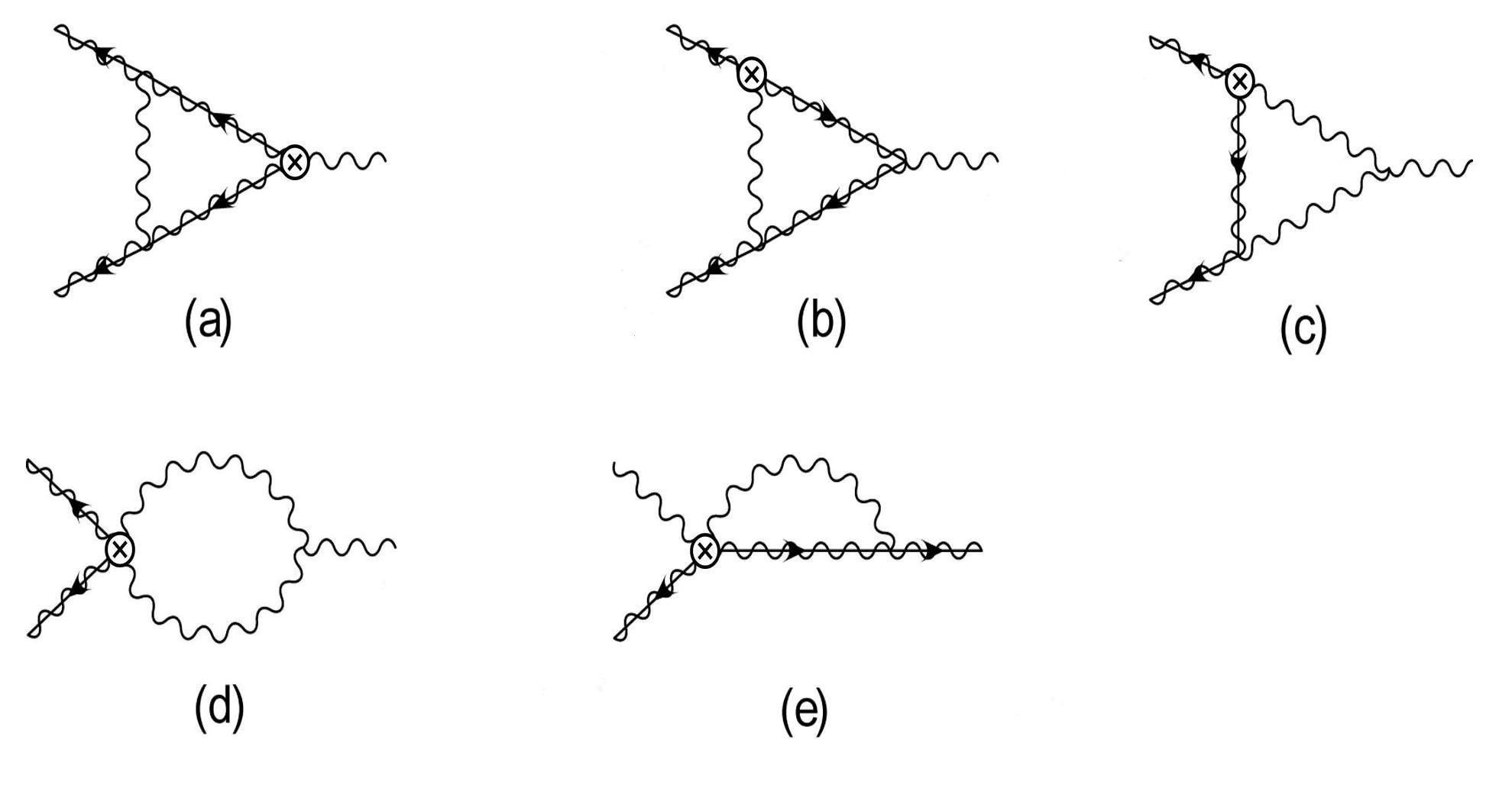}}
\caption{\label{fig1}\small 1PI three point  function diagrams with  one gauge, two gaugino lines; the crossed  circle represents the positions of 
NAC parameter $C$ associated with parameter $Y$.}
\end{figure}
\begin{figure}[ht]
\centerline{\includegraphics[width=12cm]{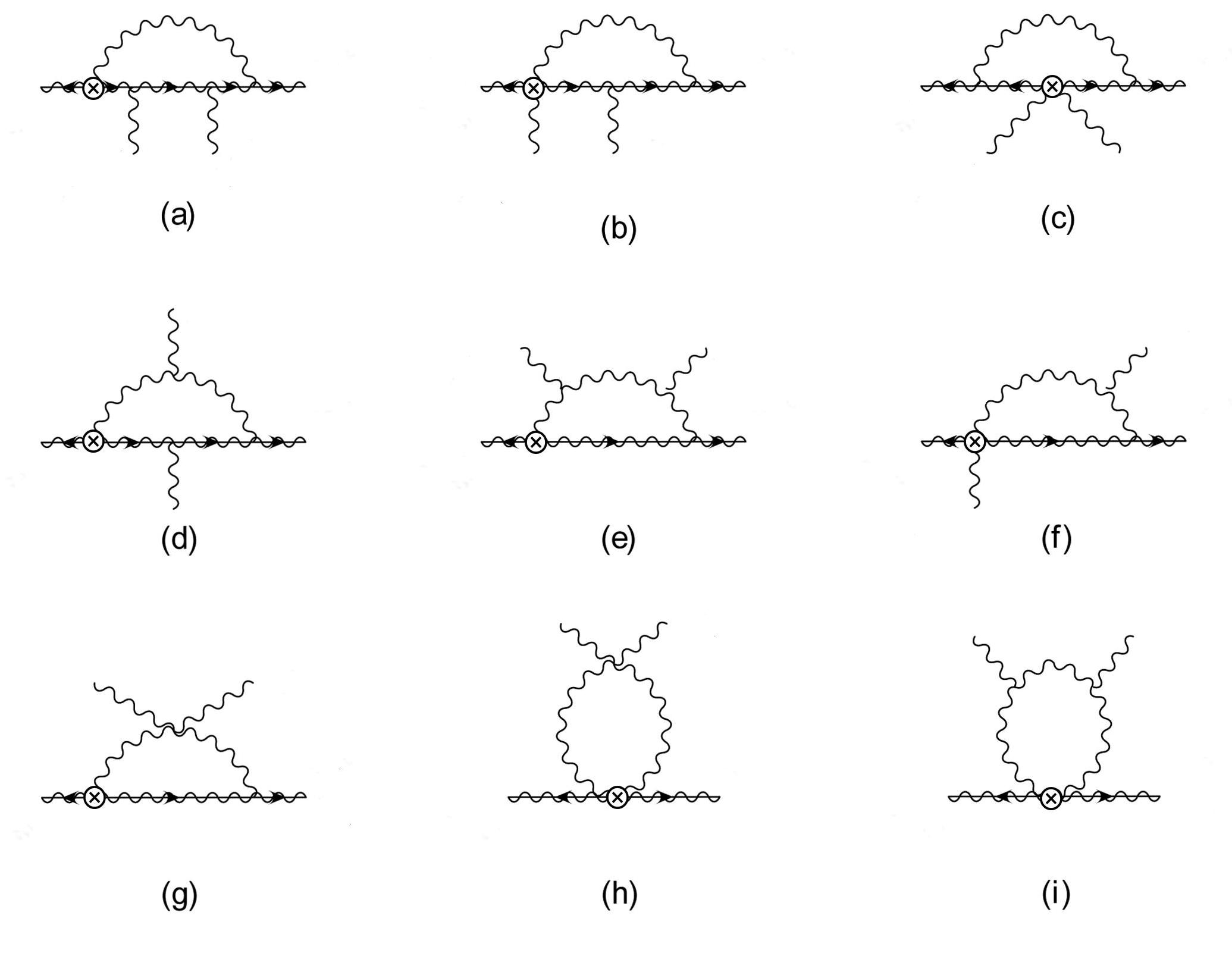}}
\caption{\label{fig1}\small 1PI four point  function diagrams with two gauge and two gaugino lines; the crossed  circle represents the positions of 
NAC parameter $C$ associated with parameter $Y$.}
\end{figure}

\end{document}